\documentclass[]{tMPH2e}
\usepackage[usenames,dvipsnames]{xcolor}
\usepackage{graphics, inputenc}
\usepackage{graphicx,amsmath,amssymb,tabularx}
\usepackage{tikz}
\usepackage{multicol}
\usepackage{epstopdf}
\usepackage{float}
\usepackage{subfigure}
\usepackage{mathtools}
\definecolor{dgreen}{rgb}{0,.5,0}
\definecolor{dred}{rgb}{.7,.0,.0}
\def\etal{{\it et al.} }

\newcommand{\masa}[1]{{\textcolor{red}{#1}} }

\DeclarePairedDelimiter\bra{\langle}{\rvert}
\DeclarePairedDelimiter\ket{\lvert}{\rangle}
\DeclarePairedDelimiterX\braket[2]{\langle}{\rangle}{#1 \delimsize\vert #2}
\def\ddroit{{\rm d}}

\begin{document}
\doi{doi}
 \issn{}
\issnp{}
\jvol{vol}
\jnum{num} \jyear{2016} \jmonth{February}

\markboth{B.~Senjean {\it et al.}}{Molecular Physics}

\articletype{Manuscript}

\title{{\itshape 
Local density approximation in site-occupation embedding theory}
}

\author{
Bruno Senjean$^1$, Masahisa Tsuchiizu$^2$, Vincent Robert$^1$, and Emmanuel Fromager$^{1{\ast}}$\thanks{$^\ast$Corresponding author.
Email: fromagere@unistra.fr 
\vspace{6pt}}
\\\vspace{6pt}  
{\em{
$^1$Laboratoire de Chimie Quantique,
Institut de Chimie, CNRS / Universit\'{e} de Strasbourg,
4 rue Blaise Pascal, 67000 Strasbourg, France\\
\vspace{0.3cm}
$^2$Department of Physics, Nagoya University, Nagoya 464-8602, Japan
}}\\\vspace{6pt}  
}

\maketitle

\begin{abstract}

Site-occupation embedding theory (SOET) is a density-functional theory
(DFT)-based method which aims at modelling strongly correlated
electrons. It is in principle exact and applicable to model
and quantum chemical Hamiltonians. The theory is presented here for the Hubbard
Hamiltonian. In contrast to conventional DFT
approaches, the site (or orbital) occupations are deduced in SOET from a
partially-interacting system consisting of one (or more) impurity
site(s) and non-interacting bath sites. The correlation
energy of the bath is then treated implicitly by means of a
site-occupation functional. In this work, we propose a simple
impurity-occupation functional approximation based on the two-level (2L)
Hubbard model which is referred to as two-level impurity local density
approximation (2L-ILDA). Results obtained on a prototypical uniform 8-site Hubbard
ring are promising. The extension of the method to larger systems and more
sophisticated model Hamiltonians is currently in progress.

\bigskip

\bigskip

\end{abstract}

\section{Introduction}\label{sec:intro}

Modelling strongly correlated electronic systems, such as materials and molecules
containing transition metals, 
remains a challenge for quantum chemists and condensed 
matter physicists. The difficulty lies in the proper description of both strong
(static) and weak (dynamical) correlation effects. Strong correlation
prevents from using a mean-field approximation as starting point for the
description of the dynamical correlation. While, in computational chemistry,
multi-reference perturbation and coupled-cluster theories (see, for example,
Refs.~\cite{andersson1992second,roca2012multiconfiguration,angeli2001introduction,angeli2002n,lyakh2011multireference})
became the state of the art, various methods have also been developed in
condensed matter physics. Let us mention, among others, quantum and variational Monte Carlo
~\cite{gros1987antiferromagnetic, sorella2005wave, neuscamman2012optimizing, zhang1997pairing, chang2010spin, yanagisawa2001ground},
the density-matrix renormalization group (DMRG) method~\cite{white1992density, white1993density}, the symmetry breaking and restoration methods
~\cite{rodriguez2013multireference, scuseria2011projected, jimenez2012projected, rodriguez2012symmetry, juillet2013exotic, tomita2004many} 
and methods based on the Gutzwiller variational approach~\cite{baeriswyl2009variational, lanata2012efficient}.
Considering that electron correlation is
local~\cite{pulay1983localizability, saebo1993local,
hampel1996local,tsuchimochi2015density}, 
alternative approaches such as the dynamical mean-field theory (DMFT)
~\cite{DMFT_correlation_limit,kotliar2004strongly,DMFT_calculations,held2007electronic,DMFT_quantum} 
decompose the whole system into a small subsystem called impurity, which
is embedded in the rest of the system referred to as the bath. 
The impurity and the bath interact with each other in some approximate
but self-consistent way.
DMFT as well as the recently proposed self-energy embedding theory
(SEET)~\cite{kananenka2015systematically,lan2015communication} 
are formulated in terms of the (frequency-dependent) one-particle Green's function.
In the latter case, the self-energy for the strongly correlated orbitals
(corresponding to the local correlation effects) is obtained from an exact diagonalisation,
which is equivalent to full configuration interaction (FCI), 
and the remaining part of the correlation (called non-local correlation)
is modelled with a self-consistent 
second-order Green's function method
(GF2)~\cite{dahlen2005self,phillips2014communication}.\\
 
Chan and coworkers~\cite{DMET_vs_DMFT, zheng2016ground} recently 
introduced a density-matrix embedding theory (DMET) which 
builds an embedded impurity model from the (frequency-independent) one-particle density matrix.
Such an approach enables the computation of ground-state properties at a
much lower computational cost. Bulik \etal~\cite{bulik2014density} proposed a simplified version of DMET,
referred to as density embedding theory (DET), which uses the
diagonal elements of the density matrix only. In DMET, the quantum impurity solver can be the FCI method, and the bath is treated in a mean-field 
approximation such as Hartree--Fock (this is called FCI-in-HF embedding
in Ref.~\cite{tsuchimochi2015density}).
In order to overcome such limitations, Tsuchimochi
\etal~\cite{tsuchimochi2015density} proposed to
use an antisymmetrized geminal power 
(AGP)~\cite{nakamura1959two, kutzelnigg1964direct, coleman1965structure} 
wave function for the bath.\\

In this paper, we explore an alternative frequency-independent embedding
scheme, called site-occupation embedding theory
(SOET),
which has been introduced recently by one of the
authors~\cite{fromager2015exact}. SOET is a
density-functional theory (DFT)-based method. More precisely, it
relies on the extension of DFT to model Hamiltonians 
~\cite{gunnarsson1986density, DFT_ModelHamiltonians, DFT_lattice,
LDA_Luttinger,carrascal2015hubbard}, which is often referred to as  
site-occupation functional theory (SOFT).
The difference
between conventional
SOFT and SOET lies in the choice of the fictitious system from which the
site occupations are obtained. In SOET, a partially-interacting system
consisting of an interacting impurity site and non-interacting bath
sites is considered, rather than a completely non-interacting system. 
The embedding of the impurity system is in principle exact and obtained
from a site-occupation correlation functional for the bath. 
{
Note that, in the standard (frequency-independent)
DFT+U method~\cite{anisimov1991band,liechtenstein1995density},
an explicit Hubbard-type
two-electron repulsion $U$ parameter is introduced in the spirit of hybrid
functionals (i.e. in a single-determinantal framework). In contrast to
DFT+U, SOET aims at 
treating explicitly (with a multideterminantal wavefunction or,
alternatively, with Green's functions) some part of the correlation
energy, namely the correlation energy of the impurity site.
One of the motivation for developing such an approach is the calculation
of local properties such as double occupations. It
can be shown (this will be presented in a separate work) that the
physical double occupation can be expressed exactly in terms of the
embedded impurity double occupation and the derivative of a
complementary site-occupation-functional correlation energy with respect
to $U$. As a result, only one single impurity site is needed to improve
the calculation of local properties. 
}\\

{
As already discussed in Ref.~\cite{fromager2015exact}, SOET can in
principle be extended to quantum chemical Hamiltonians. The purpose of
such an extension would be the formulation of a rigorous double-counting-free
multiconfiguration DFT in the orbital space. A formal analogy
can actually be made between SOET and the more standard
multideterminantal extension of DFT based on range
separation~\cite{savinbook, leininger1997combining, pollet2002combining,
srDFT}. While, in the latter case, the basic variable is the electron
density (in real space) and the separation of correlation effects is
based on the range of these effects, SOET would rely, in quantum
chemistry, on the
orbital occupations. Correlation effects would then be separated in the
orbital space, like in conventional multiconfigurational methods, thus
leading to a multiconfigurational DFT-type method. The so-called active orbitals
would then play the role of "impurity sites". Work on the orbital
dependence~\cite{fromager2015exact} of such a scheme is currently in
progress. In summary, the long-range interaction in range-separated DFT
is the analog of the impurity on-site repulsion in SOET for the Hubbard
model or the analog of the two-electron repulsion in the active space for quantum
chemistry.\\ 
}

So far, SOET
has been explored at the formal level only~\cite{fromager2015exact}. In
particular, no explicit site-occupation approximate functional was
derived. In
this work, we propose a simple local density approximation (LDA) based
on the Hubbard dimer.        
The paper is organized as follows: After a brief introduction 
to SOFT (Section~\ref{subsec:SOFT}) and SOET
(Section~\ref{subsec:SOET}), the application of SOET to the two-electron
Hubbard dimer is discussed in details in
Section~\ref{subsec:Hubbard_dimer}. An impurity-occupation functional 
approximation based on the Hubbard dimer is then introduced in Section~\ref{subsec:ILDA}.
After the computational details (Section~\ref{sec:Comp_det}), exact and
approximate results
obtained for uniform 8-site Hubbard rings are presented in
Section~\ref{sec:Results}.
Conclusions are given in Section~\ref{sec:conclu}.

\section{Theory}\label{sec:theory}

\subsection{Site-occupation functional theory}\label{subsec:SOFT}

Let us consider the second-quantized expression for the Hubbard model Hamiltonian with an external
potential $\mathbf{v}\equiv \lbrace{v_i}\rbrace_i$,  
\begin{eqnarray}\label{eq:Hubb_hamil_ext_pot}
\hat{H}=\hat{\mathcal{T}}+\hat{U}+ 
\sum_iv_i\hat{n}_{i},
\nonumber\\
\hat{\mathcal{T}}=
{-t}\sum_{<i,
j>,\sigma}\hat{a}^\dagger_{i\sigma}\hat{a}_{j\sigma},\nonumber\\
\hat{U}=
{U}\sum_i
\hat{n}_{i\uparrow}\hat{n}_{i\downarrow},
\end{eqnarray}
where $t>0$ is the hopping integral, $U>0$ is the on-site two-electron
repulsion parameter,
$\hat{n}_{i\sigma}=\hat{a}^\dagger_{i\sigma}\hat{a}_{i\sigma}$, and
$\sigma=\uparrow,\downarrow$. The site-occupation operator equals 
$\hat{n}_{i}=\hat{n}_{i\uparrow}+\hat{n}_{i\downarrow}$ and, for a given
wave function $\Psi$, the occupation of site $i$ is defined as $n_i=\langle
\Psi\vert \hat{n}_{i}\vert\Psi\rangle$.
The Hohenberg--Kohn (HK) theorem can be adapted to the Hamiltonian in Equation~(\ref{eq:Hubb_hamil_ext_pot}), i.e. there is a one-to-one correspondence
between the external potential $\mathbf{v}$,
up to a constant, and the ground-state site occupations
$\mathbf{n}\equiv\lbrace
n_i\rbrace_i${~\cite{chayes1985density,gunnarsson1986density}}. 
Consequently, for a given potential $\mathbf{v}$, 
the exact ground-state
energy $E(\mathbf{v})$ can be obtained from the following variational principle,
\begin{eqnarray}\label{eq:VP_soft}
\displaystyle E(\mathbf{v})=\underset{\mathbf{n}}{\rm min}\Big\lbrace F(\mathbf{n})+
(\mathbf{v}\vert \mathbf{n})\Big\rbrace,
\end{eqnarray}
where 
$
(\mathbf{v}\vert \mathbf{n})=
\sum_i{v_i}\,{n}_i$. 
In this approach, which is referred to as 
the site-occupation functional theory (SOFT)
in the following, the analog of the universal HK functional is a $t$- and
$U$-dependent functional of
the site occupations that can be written within the Levy--Lieb (LL)
constrained-search formalism as
\begin{eqnarray}\label{eq:fully_interacting_LL_fun}
F(\mathbf{n})=
\underset{\Psi\rightarrow
\mathbf{n}}{\rm
min}\left\lbrace \langle\Psi\vert\hat{\mathcal{T}}+\hat{U}\vert\Psi\rangle\right\rbrace.
\end{eqnarray}
Note that the site occupations are a collection of occupation numbers whereas 
the electron density, which is the basic variable in conventional DFT, is a continuous function of space
coordinates. Strictly speaking, one should refer to $F(\mathbf{n})$ as a LL
function rather than functional. Nevertheless, "functional" is often
used in the literature~\cite{DFT_ModelHamiltonians}, thus emphasizing on
the formal analogies between SOFT and DFT. We will do the same in the
rest of this paper. Note that the LL functional can be rewritten as a
Legendre--Fenchel transform~\cite{LFTransform-Lieb,fromager2015exact}:
\begin{equation}\label{LF_fully-interacting}
F(\mathbf{n}) = \underset{\mathbf{v}}{\text{sup}} \Big\{
{E}(\mathbf{v}) -(\mathbf{v}\vert \mathbf{n})  \Big\}.
\end{equation}
The maximising potential provides the ground-state wave function with
site occupations $\mathbf{n}$. As shown in
Sec.~\ref{subsec:Hubbard_dimer}, this expression is very convenient for
computing exact correlation energies in simple systems like the Hubbard
dimer.\\
\\
The Kohn--Sham (KS) formulation of
SOFT (KS--SOFT) is obtained from the following partitioning,
\begin{eqnarray}\label{eq:HxcSOfun_def}
F(\mathbf{n})=\mathcal{T}_{\rm s}(\mathbf{n})+E_{\rm Hxc}(\mathbf{n}),
\end{eqnarray} where the non-interacting kinetic energy functional is
defined in analogy with KS--DFT as
\begin{eqnarray}\label{eq:Ts(n)}
\mathcal{T}_{\rm s}(\mathbf{n})=
\underset{\Psi\rightarrow
\mathbf{n}}{\rm
min}\left\lbrace
\langle\Psi\vert\hat{\mathcal{T}}\vert\Psi\rangle\right\rbrace,
\end{eqnarray}
and $E_{\rm Hxc}(\mathbf{n})$ is the Hartree-exchange-correlation (Hxc)
site-occupation functional, which can be decomposed into a
Hartree-exchange and a correlation part as
follows~\cite{DFT_ModelHamiltonians},
\begin{eqnarray}\label{eq:Hxc_decomp}
E_{\rm Hxc} (\mathbf{n}) & = & E_{\rm Hx}(\mathbf{n}) + E_{\rm c}(\mathbf{n}),
\end{eqnarray}
where 
\begin{eqnarray}\label{eq:EHx}
E_{\rm Hx}(\mathbf{n}) & = & \dfrac{U}{4} \sum_i n_i^2.
\end{eqnarray}
The exact expression for the correlation functional is unknown. 
At the LDA level, which is exact for the uniform
Hubbard model, the site-occupation correlation functional is written as  
\begin{eqnarray}\label{eq:Ec_uniform}
{E}^{\rm LDA}_{\rm c}({\bf n})=\sum_i
e_{\rm c}(n_i),
\end{eqnarray}
where an approximate expression for the per-site correlation energy
$e_{\rm c}(n)$ can be obtained, for example, from the Bethe-ansatz (BA) solution,
thus leading to the BALDA functional of Capelle and
coworkers~\cite{LDA_Luttinger}. More on practical KS-SOFT calculations
can be found in Refs.~\cite{DFT_ModelHamiltonians,carrascal2015hubbard} and the references
therein.   


\subsection{Site-occupation embedding theory}\label{subsec:SOET}

In the spirit of 
DMET \cite{DMET_vs_DMFT,bulik2014density}, 
Fromager~\cite{fromager2015exact} proposed recently to treat the
two-electron repulsion $U$ explicitly on a limited number of 
sites called \textit{impurities} while the remaining sites, 
which play the role of a \textit{bath}, are treated within SOFT, hence
this theory is named site-occupation embedding theory
(SOET).
In the
following, we will consider only one impurity site labelled as $j_0$.
Let us stress that SOET is in
principle exact and that the ground-state energy should not depend on the number of impurity
sites if exact complementary functionals are
used~\cite{fromager2015exact}.
The exact embedding relies on the following 
partitioning of the LL functional,
\begin{eqnarray}\label{eq:LL_SOET}
F(\mathbf{n}) = F^{\rm imp}(\mathbf{n}) + 
\overline{E}^{\rm bath}_{\rm Hxc}(\mathbf{n}) ,
\end{eqnarray}
where 
\begin{eqnarray}\label{eq:LL_imp}
F^{\rm imp}(\mathbf{n}) = \underset{\Psi\rightarrow
\mathbf{n}}{\rm
min}\Big\{ \langle\Psi\vert\hat{\mathcal{T}}
+U\hat{n}_{j_0\uparrow}\hat{n}_{j_0\downarrow} \vert\Psi\rangle\Big\}.
\end{eqnarray}
Note that the latter functional can be rewritten as a Legendre--Fenchel
transform~\cite{fromager2015exact},
\begin{equation}\label{LF1}
F^{\rm imp}(\mathbf{n}) = \underset{\mathbf{v}}{\text{sup}} \Big\{
\mathcal{E}^{\rm imp}(\mathbf{v}) -(\mathbf{v}\vert \mathbf{n})  \Big\},
\end{equation}
where
\begin{eqnarray}\label{eq:variational_energy_in_LF_trans}
\mathcal{E}^{\rm imp}(\mathbf{v})= \underset{\Psi}{\rm
min}\left\lbrace \langle\Psi\vert\hat{\mathcal{T}}+U\hat{n}_{j_0\uparrow}\hat{n}_{j_0\downarrow} \vert\Psi\rangle 
+ (\mathbf{v}\vert \mathbf{n}^{\Psi}) \right\rbrace,
\end{eqnarray}
$
(\mathbf{v}\vert \mathbf{n}^{\Psi})=
\langle \Psi \vert \sum_i{v_i}\,\hat{n}_i \vert \Psi
\rangle=\sum_i{v_i}\,{n}^{\Psi}_i$, 
and $\mathbf{n}^{\Psi}$ denotes the on-site occupations for the trial
wave function $\Psi$. Let us stress that $\mathcal{E}^{\rm
imp}(\mathbf{v})$ is a (non-physical) auxiliary $N$-electron energy. The maximising potential in Equation~(\ref{LF1}), which we will refer to as embedding
potential in the following, ensures that the fictitious
embedded impurity system has the site occupations $\mathbf{n}$.\\
\\
Returning to the decomposition in Equation~(\ref{eq:LL_SOET}), the complementary site-occupation functional for the bath
$\overline{E}^{\rm bath}_{\rm Hxc}(\mathbf{n})$ depends on $t$ and $U$
only. It is, in a sense, universal as it does not depend on the external
potential. In analogy with Equations~(\ref{eq:Hxc_decomp}) and
(\ref{eq:EHx}), we decompose the bath functional as follows,
\begin{eqnarray}\label{eq:Hxc_Hx+c_bath}
\overline{E}^{\rm bath}_{\rm Hxc}(\mathbf{n})  & = & 
\overline{E}^{\rm bath}_{\rm Hx}(\mathbf{n})  + \overline{E}^{\rm
bath}_{\rm c}(\mathbf{n}), 
\end{eqnarray}
where
\begin{eqnarray}\label{eq:bath-Hx=_}
\overline{E}^{\rm bath}_{\rm Hx}(\mathbf{n}) = \dfrac{U}{4} \sum_{i \neq
j_0} n_i^2 . 
\end{eqnarray}
The correlation functional for the bath $\overline{E}^{\rm bath}_{\rm
c}(\mathbf{n})$ can be connected with the standard correlation
functional in Equation~(\ref{eq:Hxc_decomp}) by considering the KS {partitioning}
of the embedded impurity LL functional (see Equation~(\ref{eq:LL_imp})),
\begin{eqnarray}\label{eq:Fimp_Ts+Eimp}
F^{\rm imp}(\mathbf{n}) = \mathcal{T}_s(\mathbf{n}) + E^{\rm imp}_{\rm Hxc}(\mathbf{n}),
\end{eqnarray}
where the Hxc energy of the embedded impurity equals  
\begin{eqnarray}\label{eq:Hxc_Hx+c_imp}
E^{\rm imp}_{\rm Hxc}(\mathbf{n}) & = & E^{\rm imp}_{\rm Hx}(\mathbf{n})
+ E^{\rm imp}_{\rm c}(\mathbf{n}),
\end{eqnarray}
with \begin{eqnarray} \label{eq:EHximp}
E^{\rm imp}_{\rm Hx}(\mathbf{n})  & = & \dfrac{U}{4}n_{j_0}^2.
\end{eqnarray}
Combining Equations~(\ref{eq:HxcSOfun_def}), (\ref{eq:LL_SOET}) and
(\ref{eq:Fimp_Ts+Eimp}) leads to the final expression:
\begin{eqnarray}\label{eq:Ebath-c_E-c-Eimp-c}
\overline{E}^{\rm bath}_{\rm c}(\mathbf{n}) = E_{\rm c} (\mathbf{n}) - E^{\rm imp}_{\rm c}(\mathbf{n}).
\end{eqnarray}
Therefore, in order to perform practical SOET calculations, it is
necessary to develop approximations to the impurity correlation
functional $E^{\rm imp}_{\rm c}(\mathbf{n})$, in addition to the
conventional (fully-interacting) correlation functional. For the latter,
the BALDA functional~\cite{LDA_Luttinger} could be
used. This choice will be discussed in a separate work. In this paper,
we focus on approximate impurity correlation functionals that depend
only on the impurity occupation.\\ 

Returning to the exact theory, it comes from Equations~(\ref{eq:VP_soft}),
(\ref{eq:LL_SOET}) and (\ref{eq:LL_imp})
that, for any normalised trial wave function $\Psi$, 
\begin{eqnarray}\label{eq:variational_principle_impurity}
\langle\Psi\vert\hat{\mathcal{T}}+U\hat{n}_{j_0\uparrow}\hat{n}_{j_0\downarrow} \vert\Psi\rangle 
+   \overline{E}^{\rm bath}_{\rm Hxc}(\mathbf{n}^{\Psi}) + (\mathbf{v}\vert \mathbf{n}^{\Psi})
&\geq & F^{\rm imp}(\mathbf{n}^{\Psi}) + \overline{E}^{\rm bath}_{\rm Hxc}(\mathbf{n}^{\Psi})
 + (\mathbf{v}\vert \mathbf{n}^{\Psi})
\nonumber\\
& \geq & E(\mathbf{v}),
\end{eqnarray}
thus leading to the exact variational expression for the ground-state
energy:
\begin{eqnarray}\label{eq:variational_energy}
E(\mathbf{v}) &=& \underset{\Psi}{\rm
min}\left\lbrace 
\langle\Psi\vert\hat{\mathcal{T}}+U\hat{n}_{j_0\uparrow}\hat{n}_{j_0\downarrow} \vert\Psi\rangle 
+   \overline{E}^{\rm bath}_{\rm Hxc}(\mathbf{n}^{\Psi}) + (\mathbf{v}\vert \mathbf{n}^{\Psi})
 \right\rbrace
\nonumber\\
&=&
\langle\Psi^{\rm
imp}\vert\hat{\mathcal{T}}+U\hat{n}_{j_0\uparrow}\hat{n}_{j_0\downarrow}
\vert\Psi^{\rm imp}\rangle 
+   \overline{E}^{\rm bath}_{\rm Hxc}(\mathbf{n}^{\Psi^{\rm imp}}) +
(\mathbf{v}\vert \mathbf{n}^{\Psi^{\rm imp}})
.
\end{eqnarray}
The minimising wave function $\Psi^{\rm imp}$ fulfills the following self-consistent Equation~\cite{fromager2015exact}:
\begin{equation}\label{eq:self-consistent-SOET}
\displaystyle \left( \hat{\mathcal{T}} + U \hat{n}_{j_0\uparrow}\hat{n}_{j_0\downarrow}
 + \displaystyle \sum_{i} \left[v_i + 
 \dfrac{\partial \overline{E}^{\rm bath}_{\rm Hxc}(\mathbf{n}^{\Psi^{\rm imp}})}{\partial n_i} \right]
 \hat{n}_i \right) \ket{\Psi^{\rm imp}}  = \mathcal{E}^{\rm imp} \ket{\Psi^{\rm imp}} ,
\end{equation}
where $\mathcal{E}^{\rm imp}$ is the energy of the 
embedded impurity system. Consequently, the exact embedding potential
can be expressed as follows, 
\begin{eqnarray}\label{eq:embpot}
v^{\rm emb}_i = v_i + \dfrac{\partial \overline{E}^{\rm bath}_{\rm Hxc}(\mathbf{n}^{\Psi^{\rm imp}})}{\partial n_i}.
\end{eqnarray}
The latter ensures that the physical (fully-interacting) and fictitious
embedded impurity systems have exactly the same site occupations. 


\subsection{Application to the Hubbard dimer}\label{subsec:Hubbard_dimer}

\subsubsection{Notations and energy expressions}

In the spirit of a recent work by Carrascal~\etal~\cite{carrascal2015hubbard}, we propose in
this section to apply SOET to the two-electron Hubbard dimer. For that
purpose, we consider the general Hamiltonian expression:
\begin{eqnarray}
\hat{\mathcal{H}}&=&-t\sum_\sigma\Big(
\hat{a}^\dagger_{0\sigma}\hat{a}_{1\sigma}+
\hat{a}^\dagger_{1\sigma}\hat{a}_{0\sigma}
\Big)
+U_0\hat{n}_{0\uparrow}\hat{n}_{0\downarrow}
\nonumber\\
&&+U_1\hat{n}_{1\uparrow}\hat{n}_{1\downarrow}
+v_0\hat{n}_{0}+v_1\hat{n}_{1}.
\end{eqnarray}
Let us mention that the uniform case ($n_0=n_1=1$) has already been solved in
Ref.~\cite{fromager2015exact}. For simplicity, we will
assume that 
\begin{eqnarray}\label{eq:pot_cond}
v_0+v_1=0.
\end{eqnarray}
Note that the latter condition is fulfilled by any potential once it has
been shifted by $-(v_0+v_1)/2$. 
The physical and non-interacting KS systems are obtained when 
$U_0=U_1=U$ and $U_0=U_1=0$, respectively. The embedded impurity
system, where the impurity site is labelled as 0, corresponds to $U_0=U$
and $U_1=0$.\\ 
Since we are interested in the
singlet ground state only, the matrix representation of the Hamiltonian can
be reduced to the basis of the "doubly-occupied site" states $\vert
D_i\rangle=\hat{a}^\dagger_{i\uparrow}\hat{a}^\dagger_{i\downarrow}\vert{\rm
vac}\rangle$ with $i=0$ or $1$, and $\vert
S\rangle=1/\sqrt{2}(\hat{a}^\dagger_{0\uparrow}\hat{a}^\dagger_{1\downarrow}
-\hat{a}^\dagger_{0\downarrow}\hat{a}^\dagger_{1\uparrow})\vert
{\rm vac}\rangle$ that corresponds to singly-occupied sites, thus
leading to~\cite{fromager2015exact}:       
\begin{eqnarray}\label{Hmatrix}
\left[ \hat{\mathcal{H}} \right]
&=&
\left[
\begin{array}{c c c}
U_0-\delta v & 0 & -\sqrt{2}t\\
0 & U_1+\delta v & -\sqrt{2}t\\ 
-\sqrt{2}t & -\sqrt{2}t & 0 
\end{array}
\right]
,
\end{eqnarray}
where $\delta v=v_1-v_0$. For a physical system
($U_0=U_1=U$), the exact 
ground-state energy can be expressed analytically as
follows~\cite{carrascal2015hubbard}, 
\begin{eqnarray}\label{eq:physical_energy_dimer}
E(U,\delta v)=\dfrac{4t}{3}\left(u-w\,{\rm sin}\left(\theta +\dfrac{\pi}{6}\right)\right),
\end{eqnarray}
where 
\begin{eqnarray}
u=U/2t,
\\ 
w=\sqrt{3(1+\nu^2)+u^2},
\\
\nu=\delta v/2t,
\end{eqnarray}
and
\begin{eqnarray}\label{eq:energy_phys_HUbb_end}
{\rm cos}(3\theta)=\left(9(\nu^2-1/2)-u^2\right)u/w^3.
\end{eqnarray}
The expression for the exact ground-state energy
$\mathcal{E}(U_0,U_1,\delta v)$ of the general
Hamiltonian matrix in Equation~(\ref{Hmatrix}) is then obtained
straightforwardly by considering an effective physical system such that 
\begin{eqnarray}
U_{\rm eff}-\delta v_{\rm eff}=U_0-\delta v,
\nonumber\\
U_{\rm eff}+\delta v_{\rm eff}=U_1+\delta v,
\end{eqnarray} 
or, equivalently,
\begin{eqnarray}
U_{\rm eff}&=& \dfrac{U_0+U_1}{2},\\
\delta v_{\rm eff}&=& \delta v + \dfrac{U_1-U_0}{2}.
\end{eqnarray} 
As a result, we obtain the final expression:
\begin{eqnarray}
\mathcal{E}(U_0,U_1,\delta v)=E\Big((U_0+U_1)/2,\delta v
+(U_1-U_0)/2\Big).
\end{eqnarray}
In the particular case of the embedded impurity system, the energy
becomes
\begin{eqnarray}\label{eq:scaling_full_imp}
\mathcal{E}^{\rm imp}(U,\delta v)=\mathcal{E}(U,0,\delta v)=E\big(U/2,\delta v-U/2\big).
\end{eqnarray}

For conveniency, the occupation of site 0 will be denoted as $n_0=n$.
Consequently, $n_1=2-n$ since the number of electrons is held
constant and equal to 2. Therefore, in this simple system, the site occupations reduce to a
single number $n$ which can vary from 0 to 2.

\subsubsection{Non-interacting and physical systems}

As shown in Ref.~\cite{carrascal2015hubbard}, the non-interacting kinetic energy and the KS potential can be expressed
analytically as follows,
\begin{eqnarray}\label{eq:Ts}
\mathcal{T}_{\rm s}(n) = -2t\sqrt{n(2-n)} ,
\end{eqnarray}
and
\begin{eqnarray}\label{eq:_kspot}
\delta v^{\rm KS}(n) 
=\frac{\partial \mathcal{T}_{\rm s}(n)}{\partial n}
= \dfrac{2t(n-1)}{\sqrt{n(2-n)}}.
\end{eqnarray}
On the other hand, there is no simple analytical expression for the
fully-interacting LL functional. 
Nevertheless, the latter can be
computed "exactly" by means of the Legendre--Fenchel transform in
Equation~(\ref{LF_fully-interacting}). Using
Equation~(\ref{eq:pot_cond}) leads to the final expression: 
\begin{eqnarray}\label{full_LF_transf}
F(U,n)&=&
\underset{\delta v
}{\rm sup}\Big\lbrace 
{E}(U,\delta v)+\delta v \times (n-1)
\Big\rbrace
\\
\nonumber
&=&
{E}\Big(U,\delta v(U,n)\Big)+\delta v (U,n)\times (n-1)
.
\end{eqnarray}
Note that the maximising potential $\delta v(U,n)$ ensures that the physical
ground-state occupation of site 0 equals $n$. It then becomes possible to compute the exact correlation energy
as follows,
\begin{eqnarray}\label{eq:corr_fun_dimer}
E_c(U,n)=F(U,n)-\mathcal{T}_{\rm s}(n)-E_{\rm Hx}(U,n),
\end{eqnarray}
where
\begin{eqnarray}\label{eq:EHX(U,n)}
E_{\rm Hx}(U,n)&=&\dfrac{U}{4}\left(n^2_0+n^2_1\right)
\nonumber\\
&=&U\Big(1-n(2-n)/2\Big).
\end{eqnarray}
Interestingly, as readily seen in
Equations~(\ref{eq:physical_energy_dimer})--(\ref{eq:energy_phys_HUbb_end}),
the physical energy is an even function of the potential,
\begin{eqnarray}\label{eq:evenfunction_E}
E(U,\delta v) = E(U,-\delta v).
\end{eqnarray}
As a result,  the
substitution $n\rightarrow 2-n$ in Equation~(\ref{full_LF_transf}) leads
to
\begin{eqnarray}
F(U,2-n)&=&
\underset{\delta v
}{\rm sup}\Big\lbrace 
{E}(U,-\delta v)-\delta v \times (n-1)
\Big\rbrace
\\
\nonumber
&=&
{E}\Big(U,-\delta v(U,2-n)\Big)-\delta v (U,2-n)\times (n-1)
,
\end{eqnarray}
which gives, by comparison with Equation~(\ref{full_LF_transf}), the hole/particle symmetry relation for the functional,
\begin{eqnarray}\label{eq:F(2-n)=F(n)}
F(U,2-n)=F(U,n),
\end{eqnarray}
and the corresponding antisymmetry relation for the physical
potential: 
\begin{eqnarray}\label{eq:v(2-n)=-v(n)}
\delta v(U,2-n) = -\delta v(U , n). 
\end{eqnarray}
Since both the non-interacting kinetic energy and the Hx energy (see
Equations~(\ref{eq:Ts}) and (\ref{eq:EHX(U,n)})) exhibit the same
symmetry, 
\begin{eqnarray}
\mathcal{T}_{\rm s}(2-n)&=&\mathcal{T}_{\rm s}(n),
\\
E_{\rm Hx}(U,2-n)&=&E_{\rm Hx}(U,n)\label{eq:hp_sym_Hx},
\end{eqnarray}
we conclude that 
\begin{eqnarray}\label{eq:Ec(2-n)=Ec(n)}
E_{\rm c}(U,2-n) = E_{\rm c}(U,n).
\end{eqnarray}
In addition, we see from Equation~(\ref{eq:_kspot}) that, for the KS
system, the following
hole/particle antisymmetry relation is fulfilled:
\begin{eqnarray}\label{eq:anti_vKS}
\delta v^{\rm KS}(2-n)=-\delta v^{\rm KS}(n).
\end{eqnarray}
As a result, the Hxc potential $\delta v_{\rm Hxc}(U,n) = \delta v^{\rm
KS}(n) - \delta v(U,n)$ fulfills the same antisymmetry relation:
\begin{eqnarray}\label{eq:hp_antisym_Hxc}
\delta v_{\rm Hxc} (U,2 - n) = - \delta v_{\rm Hxc} (U,n).
\end{eqnarray}
Note that this antisymmetric behavior of the Hxc potential is directly
connected to the hole/particle symmetry of the Hxc energy. This appears
clearly from the following alternative derivation: according to Equation~(\ref{full_LF_transf}), we have for any trial
site occupation 
$\rho$,
\begin{eqnarray}
F(U,\rho)\geq E\Big(U,\delta v(U,n)\Big)+\delta v(U,n) \times (\rho-1),
\end{eqnarray}  
thus leading to
\begin{eqnarray}\label{full_VP_dens}
E\big(U,\delta v(U,n)\big)&=&
\underset{\rho
}{\rm min}\Big\lbrace 
F(U,\rho)-\delta v(U,n)\times (\rho-1)
\Big\rbrace
,
\end{eqnarray}
and, consequently, to
\begin{eqnarray}
\delta v(U,n)=\dfrac{\partial F(U,n)}{\partial n}.
\end{eqnarray}
We obtain from Equation~(\ref{eq:_kspot}) the expected expression for the Hxc potential:
\begin{eqnarray}\label{eq:vHxc=-dEHxc/dn}
\delta v_{\rm Hxc}(U,n) &=& -
\dfrac{\partial E_{\rm Hxc}(U,n)}{\partial n}
\nonumber\\
&=&
\delta v_{\rm Hx}(U,n)-
\dfrac{\partial E_{\rm c}(U,n)}{\partial n},
\end{eqnarray}
where $\delta v_{\rm Hx}(U,n)=U(1-n)$. Thus, we recover
Equation~(\ref{eq:hp_antisym_Hxc}) from 
Equations~(\ref{eq:hp_sym_Hx}) and (\ref{eq:Ec(2-n)=Ec(n)}).
Note finally the minus sign on the right-hand side of
Equation~(\ref{eq:vHxc=-dEHxc/dn}) which is simply related to the fact
that $\delta v_{\rm Hxc}(U,n)$ is the difference between the Hxc potential
values on sites 1 and 0:
\begin{eqnarray}
\left.\Bigg(\dfrac{\partial E_{\rm Hxc} (\mathbf{n})
}{\partial n_1}
-
\dfrac{\partial E_{\rm Hxc} (\mathbf{n})
}{\partial n_0}\Bigg)\right|_{n_0=n,n_1=2-n}
&=&-\dfrac{\partial}{\partial n}\Big[E_{\rm Hxc} (n,2-n)\Big]
\nonumber\\
&=&
\delta v_{\rm Hxc}(U,n),
\end{eqnarray}
since, by construction, $E_{\rm Hxc} (n,2-n)=E_{\rm Hxc}(U,n)$.

\subsubsection{Embedded impurity system}

In analogy with Equation~(\ref{full_LF_transf}), the Legendre--Fenchel transform for
the embedded impurity system (see Equation~(\ref{LF1})) can be rewritten as 
\begin{eqnarray}\label{imp_LF_transf}
F^{\rm imp}(U,n)&=&
\underset{\delta v
}{\rm sup}\Big\lbrace
\mathcal{E}^{\rm imp}(U,\delta v)+\delta v\times(n-1)
\Big\rbrace
,
\end{eqnarray}
where the maximising potential $\delta v^{\rm emb}(U,n)$
ensures that the ground-state impurity occupation in the embedded impurity
system equals $n$. Using Equations~(\ref{eq:scaling_full_imp}) and
(\ref{full_LF_transf}) leads to the following   
scaling/shifting relations for the functional,
\begin{eqnarray}\label{scaling_relations_Fimp}
F^{\rm imp}(U,n)=F(U/2,n)+\dfrac{U}{2}(n-1),
\end{eqnarray}
and for the maximising potential,
\begin{eqnarray}\label{eq:vemb=vU/2+U/2}
\delta v^{\rm emb}(U,n)=\delta v(U/2,n)+U/2.
\end{eqnarray}
Consequently, the embedded impurity Hxc potential $
\delta v^{\rm imp}_{\rm Hxc}(U,n) = \delta v^{\rm KS}(n)-\delta v^{\rm emb}(U,n)
$ can
be expressed as
\begin{eqnarray}\label{eq:VHxcimp_vHxcfull_scaling}
\delta v^{\rm imp}_{\rm Hxc}(U,n)=\delta v_{\rm Hxc}(U/2,n)-U/2.
\end{eqnarray}
Note that, according to Equation~(\ref{eq:hp_antisym_Hxc}), the following hole/particle
antisymmetry relation is fulfilled:
\begin{eqnarray}\label{eq:vHxcimp(2-n)=-vHxcimp(n) - U}
\delta v_{\rm Hxc}^{\rm imp} (U,2 - n) = - \delta v_{\rm Hxc}^{\rm imp}
(U,n) - U.
\end{eqnarray}
The correlation energy of the embedded impurity is defined as
\begin{eqnarray}
E^{\rm imp}_{\rm c}(U,n)=F^{\rm imp}(U,n)-\mathcal{T}_{\rm s}(n)-E^{\rm
imp}_{\rm
Hx}(U,n), 
\end{eqnarray}
where \begin{eqnarray}\label{eq:EHxcimp(U,n)}
E^{\rm imp}_{\rm Hx}(U,n)=\dfrac{U}{4}n^2.
\end{eqnarray}
According to Equations~(\ref{eq:corr_fun_dimer}), (\ref{eq:EHX(U,n)})
and (\ref{scaling_relations_Fimp}), we conclude that the latter
correlation energy is simply obtained
from the conventional fully-interacting correlation energy by scaling 
(by $1/2$) the $U$ parameter:
\begin{eqnarray}\label{eq:Ecimp=EcU/2}
E_{\rm c}^{\rm imp}(U,n) = E_{\rm c}(U/2,n).
\end{eqnarray}
This scaling relation provides a remarkably simple 
connection between SOET and KS-SOFT. Moreover it gives, when combined
with Equations~(\ref{eq:vHxc=-dEHxc/dn}) and
(\ref{eq:VHxcimp_vHxcfull_scaling}), the expected expression
\begin{eqnarray}
\delta v_{\rm Hxc}^{\rm imp}(U,n) &=& -\dfrac{\partial E_{\rm Hxc}^{\rm imp}(U,n)}{\partial n}
\\
&=&\delta v_{\rm Hx}^{\rm imp}(U,n)-\dfrac{\partial E_{\rm c}^{\rm
imp}(U,n)}{\partial n},
\end{eqnarray}
where $\delta v_{\rm Hx}^{\rm imp}(U,n)=-Un/2$. Note finally that,
according to Equation~(\ref{eq:Ec(2-n)=Ec(n)}), the correlation energy
of the embedded impurity fulfills the hole/particle symmetry relation:\\
\begin{eqnarray}\label{eq:Eimpc(2-n)=Eimpc(n)}
E_{\rm c}^{\rm imp}(U,2-n) = E_{\rm c}^{\rm imp}(U,n).
\end{eqnarray}
 
\begin{figure}
\centering{\includegraphics[scale=0.4]{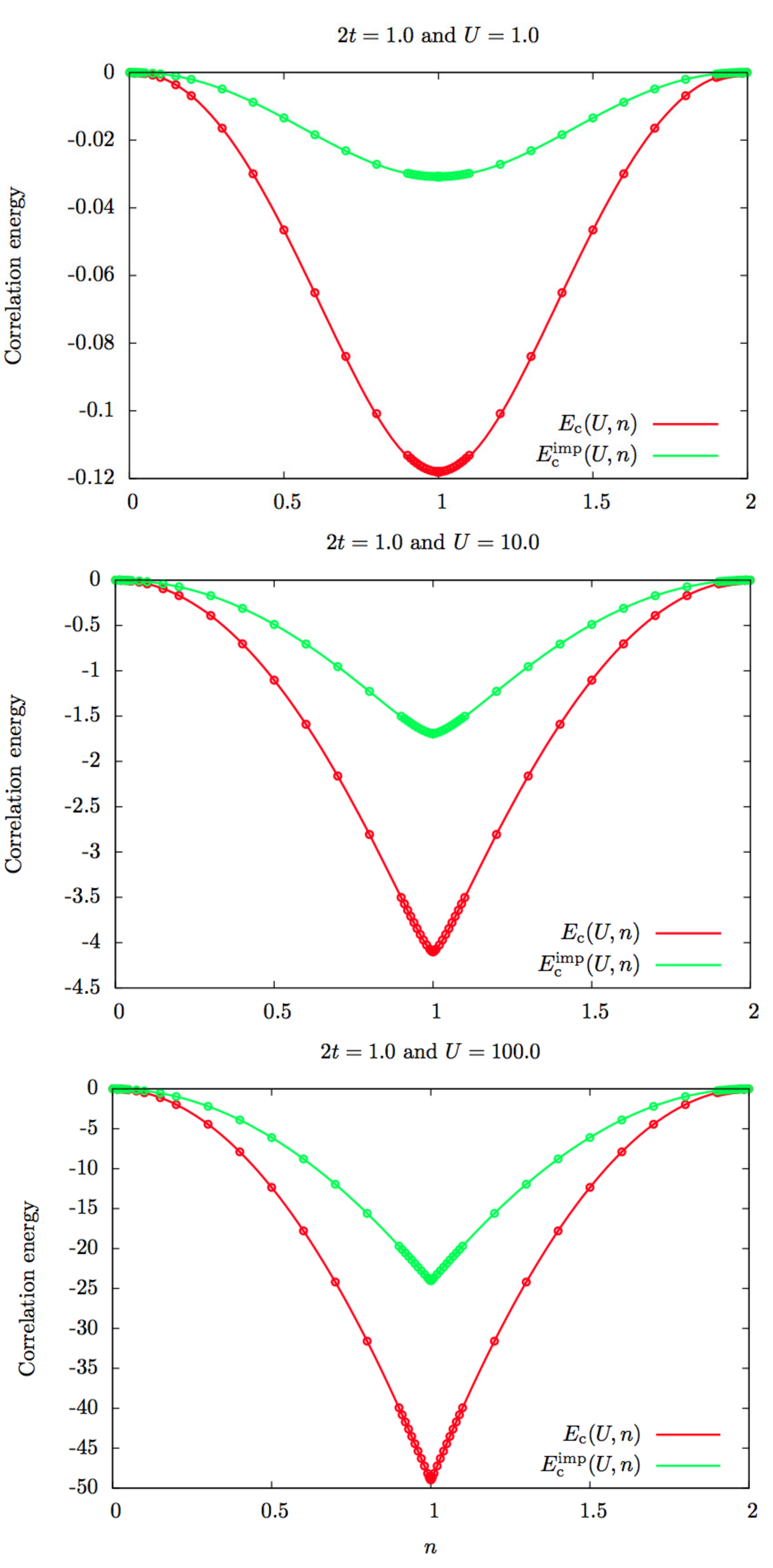}
}
\caption{Exact site-occupation functional correlation energies obtained
for the embedded impurity
and physical (fully-interacting) Hubbard dimer 
with $U = 1$ (top), $U =
10$ (middle) and $U = 100$ (bottom). 
Results obtained with the parametrisation of Carrascal \textit{et al.}
\cite{carrascal2015hubbard} are shown with open circles. See text for
further details.
}
\label{fig:Ec_Ecimp}
\end{figure}

\begin{figure}
\centering{\includegraphics[scale=0.4]{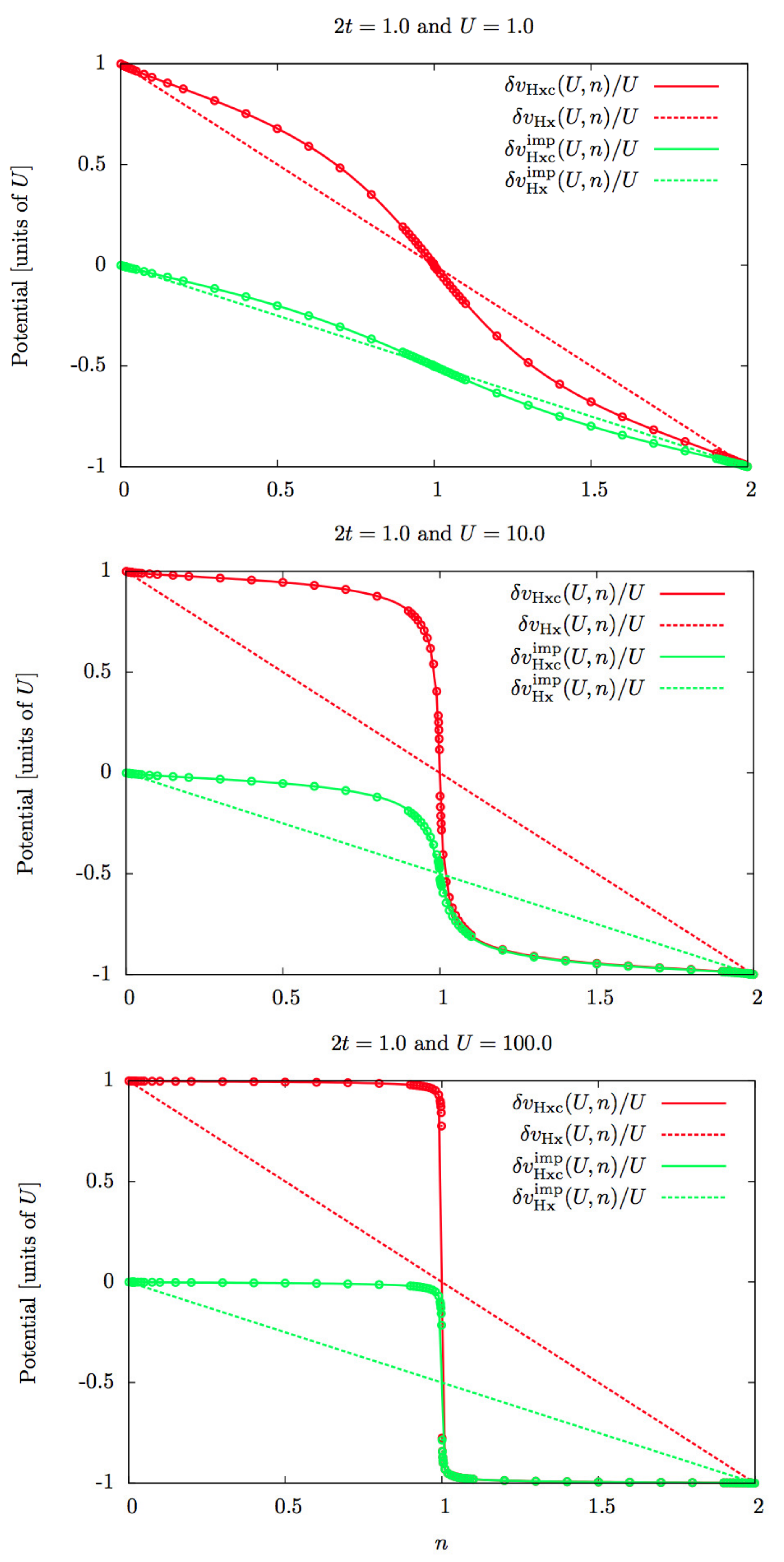}
}
\caption{Mean-field (Hx) and exact site-occupation-functional Hxc 
potentials obtained for the
embedded impurity system and the physical (fully-interacting) Hubbard dimer 
with $U = 1$ (top), $U =
10$ (middle) and $U = 100$ (bottom). 
Results obtained with the parametrisation of Carrascal \textit{et al.}
\cite{carrascal2015hubbard} are shown with open circles. See text for
further details.}
\label{fig:v_vimp}
\end{figure}

The "exact" conventional and impurity correlation site-occupation
functionals obtained by computing Legendre--Fenchel transforms are shown in Figure~\ref{fig:Ec_Ecimp}. The corresponding Hxc potentials
are plotted in Figure~\ref{fig:v_vimp}. 
Scaling as well as hole/particle symmetry and antisymmetry relations are
clearly illustrated. 
Note that the parametrisation of Carrascal \textit{et
al.}~\cite{carrascal2015hubbard}, which is also given in the Figures, provides accurate correlation energies and
potentials for all $U$ and $n$ values, as expected. In the following, we
will show how it can actually be used in practical SOET calculations.
Returning to Figure~\ref{fig:v_vimp}, the deviation from the mean-field approximation (no correlation) becomes significant when $U/t$ increases.
Interestingly, correlation effects are slightly attenuated in the
embedded impurity system because of the scaling by $1/2$ of the $U$
parameter. This is directly connected to the fact that the bath site is
non-interacting.\\

As already shown by Carrascal \textit{et
al.}~\cite{carrascal2015hubbard}, the potentials exhibit a discontinuity
at $n=1$ (i.e. when both sites are half-filled) in the strongly
correlated limit ($U/t\rightarrow+\infty$). 
A similar discontinuity can be observed in the Anderson junction model~\cite{bergfield2012bethe, liu2012accuracy, liu2015coulomb}.
It can be interpreted as follows: the impurity site with
occupation $n$ is an open system and, in order to let $n$ vary 
continuously from $0$ to $1$, we need a grand canonical ensemble
consisting of degenerate zero- and one-electron states. The latter
states are indeed degenerate (and therefore mixable) if the potential
$v_0$ on the impurity site equals zero. On the other
hand, when $n$ varies continuously from 1 to 2, the ensemble should
consist of one- and two-electron states. The latter become degenerate if
$v_0=2v_0+U$ or, equivalently, if $v_0=-U$. Since the bath is
non-interacting, the potential $v_1$ can be set to zero for any $n$
values, thus leading to $\delta v^{\rm emb}=0$ when $0<n<1$ and
$\delta v^{\rm emb}=U$
when $1<n<2$. For a non-interacting system, the potential $\delta v^{\rm
KS}$ should be zero
for any $n$ so that zero-, one- and two-electron states can be mixed, as
illustrated in Figure 9 (right panel) of
Ref.~\cite{carrascal2015hubbard}. As a result, we conclude that $\delta
v_{\rm Hxc}^{\rm imp}=\delta v^{\rm KS}-\delta v^{\rm emb}=0$ for
$0<n<1$ and $\delta v_{\rm Hxc}^{\rm imp}=-U$ for $1<n<2$, which is in
perfect agreement with the bottom panel of Figure~\ref{fig:v_vimp}.
For the fully-interacting Hxc potential, we can do a similar analysis:
if $0<n<1$, then $v_0=0$ (so that zero- and one-electron states can be
mixed on site 0) and $v_1=-U$ (so that one- and two-electron states can
be mixed on site 1), thus leading to $\delta v=-U$ and $\delta v_{\rm
Hxc}=\delta v^{\rm KS}-\delta v=U$. When $1<n<2$, zero- and one-electron
states should mix on site 1  while one- and two-electron states will mix
on site 0, thus leading to $v_0=-U$, $v_1=0$, $\delta v=U$ and,
consequently, $\delta v_{\rm Hxc}=-U$, which is again in agreement with the bottom
panel of 
Figure~\ref{fig:v_vimp}.\\

The Hubbard dimer is a two-level (2L) system. Therefore, in the
following, the superscript "2L" will be used for referring to the impurity
correlation functional in Equation~(\ref{eq:Ecimp=EcU/2}). In practice,
the parametrisation of Ref.~\cite{carrascal2015hubbard} can be used with
the substitution $U\rightarrow U/2$. 

\subsection{Impurity-occupation functional approximation}\label{subsec:ILDA}

In order to apply SOET to larger rings or infinite Hubbard systems,
approximations to the site-occupation-functional correlation energy 
of the embedded impurity system must be developed. Their combination with conventional
fully-interacting correlation functionals (like the BALDA of Capelle and
coworkers~\cite{LDA_Luttinger}) would give approximate site-occupation 
functionals for the bath (see
Equation~(\ref{eq:Ebath-c_E-c-Eimp-c})).\\
\\
A natural starting point is the LDA which consists
in applying SOET to the uniform Hubbard Hamiltonian. In this case, like
in the non-uniform case, the impurity correlation energy is in principle
a functional of all site
occupations (not only the impurity site occupation $n_{j_0}$). For simplicity, we
will assume that it does not vary with the bath site
occupations:
\begin{eqnarray}
E_{\rm c}^{\rm imp}(\mathbf{n}) \rightarrow E_{\rm c}^{\rm
imp}(n_{j_0}),
\end{eqnarray}
and shall refer to this approximation as {impurity} LDA (ILDA). Let us
stress that, for uniform systems, ILDA is an approximation whereas LDA
(see Equation~(\ref{eq:Ec_uniform})) is
exact. Indeed, in the uniform case ($v_i = 0$), the exact embedding
potential equals, according to Equations~(\ref{eq:Hxc_Hx+c_bath}),
(\ref{eq:Ebath-c_E-c-Eimp-c}) and (\ref{eq:embpot}),    
\begin{eqnarray}
v_i^{\rm emb} = 
\dfrac{U}{2}n_{i}(1-\delta_{ij_0})+
\dfrac{\partial E_{\rm c}(\mathbf{n})}{\partial n_i} - \dfrac{\partial E_{\rm c}^{\rm imp}(\mathbf{n})}{\partial n_i},
\end{eqnarray}
which becomes within ILDA:
\begin{eqnarray}\label{notshiftedembpot}
v_i^{\rm emb} \rightarrow  
\dfrac{U}{2}n_{i}(1-\delta_{ij_0})+
\dfrac{\partial e_{\rm c}(n_i)}{\partial n_i} - 
\dfrac{\partial E_{\rm c}^{\rm imp}(n_{j_0})}{\partial
n_{j_0}}\delta_{ij_0}
.
\end{eqnarray}
Since, in the exact theory, the occupations are equal ($n_i=n$),
we finally obtain the (approximate) ILDA potential expression by
shifting the potential in Equation~(\ref{notshiftedembpot}) by $-Un/2-\partial e_c(n)/\partial n$:
\begin{eqnarray}\label{eq:embpot_ILDA}
v_i^{\rm ILDA} =-\delta_{ij_0}\Bigg(\dfrac{U}{2}n_{j_0}+
\dfrac{\partial E_{\rm c}^{\rm imp}(n_{j_0})}{\partial
n_{j_0}}
\Bigg).  
\end{eqnarray}
Note that, by definition, the ILDA embedding potential is zero in the
bath. As shown numerically in Sec.~\ref{subsec:exact_embedding_potential}, this cannot, in general, lead
to strictly uniform site occupations. 
{Moreover, according to
Equations~(\ref{eq:Ebath-c_E-c-Eimp-c}) and (\ref{eq:variational_energy}),
it is necessary to employ a  
conventional site-occupation correlation functional such as
BALDA~\cite{LDA_Luttinger} in order to compute an approximate total
ILDA energy. The question whether, for practical purposes, BALDA should be
included into
the self-consistent calculation of the embedded wavefunction (see
Equation~(\ref{notshiftedembpot})) or not
raises naturally. Work is currently in progress in this direction and the results will be
presented in a separate paper. 
}\\

Practical ILDA calculations can be performed on a uniform Hubbard system by inserting the above 
expression into the self-consistent
Equation~(\ref{eq:self-consistent-SOET}), thus leading to
\begin{eqnarray}\label{eq:self-cons_ILDA}
\hspace{-0.7cm}\left( \displaystyle \hat{\mathcal{T}} + U \hat{n}_{j_0\uparrow}\hat{n}_{j_0\downarrow}
 - \Bigg[\dfrac{U}{2}n_{j_0}^{\Psi^{\rm ILDA}}+\dfrac{\partial E^{\rm imp}_{\rm c}(n_{j_0}^{\Psi^{\rm
ILDA}})}{\partial n_{j_0}}\Bigg] \hat{n}_{j_0} \right) 
 \ket{\Psi^{\rm ILDA}} = \mathcal{E}^{\rm ILDA}\ket{\Psi^{\rm ILDA}}.
\end{eqnarray}
In order to turn Equation~(\ref{eq:self-cons_ILDA}) into a computational
method, we need an impurity-occupation correlation functional
$E^{\rm imp}_{\rm c}(n_{j_0})$. We simply propose to use the functional
obtained for the (two-level) Hubbard dimer, which gives, according to
Equation~(\ref{eq:Ecimp=EcU/2}), 
\begin{eqnarray}\label{eq:2L-ILDA_energy_functional}
E_{\rm c}^{\rm imp}(n_{j_0})\rightarrow E_{\rm c}^{\rm 2L}(U/2, n_{j_0}).
\end{eqnarray}
This approximation will be referred to as 2L-ILDA.
For comparison, we discuss in the following the computation of an approximate Legendre--Fenchel
transform (see Equation~(\ref{LF1})) where the potential is set to zero in the bath, in
analogy with ILDA: 
\begin{eqnarray}\label{LF2}
F^{\rm imp}(\mathbf{n}) \rightarrow \underset{v_{j_0}}{\text{sup}}  \lbrace \mathcal{E}^{\rm imp}(v_{j_0}) - \displaystyle v_{j_0}n_{j_0} \rbrace.
\end{eqnarray}
The maximising potential in Equation~(\ref{LF2}) will be referred to as
impurity-optimised potential in the following.

\section{Computational details}\label{sec:Comp_det}

As a proof of concept, SOET has been applied to the uniform 8-site
Hubbard ring ($L=8$) with periodic or antiperiodic boundary conditions.
The number of electrons $N$ will always be even and will vary from 2 to 14.
In order to remove the pathological degeneracy,
we follow the standard treatment of the boundary conditions:
In the case of $N=2,6,10,14$, 
 we adopt the periodic boundary condition 
  ($\hat{a}_{L+1,\sigma}=\hat{a}_{1,\sigma}$),
and in the case of $N=4,8,12$, 
 we adopt the antiperiodic boundary condition 
  ($\hat{a}_{L+1,\sigma}=-\hat{a}_{1,\sigma}$). Legendre--Fenchel transforms in Equations~(\ref{LF1}) and (\ref{LF2})
have been implemented as well as the 2L-ILDA (see Equations~(\ref{eq:self-cons_ILDA}) and
(\ref{eq:2L-ILDA_energy_functional})). In the latter case, the parametrisation of Carrascal \textit{et
al.}~\cite{carrascal2015hubbard} (see Equations (102)--(107) in their
paper) has been used for the correlation functional of the dimer. Exact
diagonalisations (FCI) based on the Lanczos algorithm~\cite{lanczos} have been used in all
models. 
The convergence threshold was set to $10^{-6}$ for the site
occupations. Exact impurity correlation energies were computed as
follows, 
\begin{eqnarray}\label{eq:Ecimp=Fimp-Ts-EHximp_exact}
 E^{\rm imp}_{\rm c}(\mathbf{n}) = F^{\rm imp}(\mathbf{n}) -
\mathcal{T}_s(\mathbf{n}) - \dfrac{U}{4}n^2_{j_0}.
\end{eqnarray}
Note that the embedded impurity Legendre--Fenchel transform in
Equation~(\ref{LF1}) reduces to the non-interacting kinetic energy when
$U=0$. The hopping parameter was set to $t = 1/2$.
{Therefore, energies will be computed per unit of $2t$}. 
{Usual $U = 1$ and $U = 10$ values are considered. The
extreme $U = 100$ value is also considered in order to illustrate the discontinuity of the exchange-correlation potential at half-filling in the strongly correlated limit.} 

\section{Results and discussion}\label{sec:Results}

\subsection{Exact embedding potentials}\label{subsec:exact_embedding_potential}

\begin{figure}
\centering{
\includegraphics[scale=0.4]{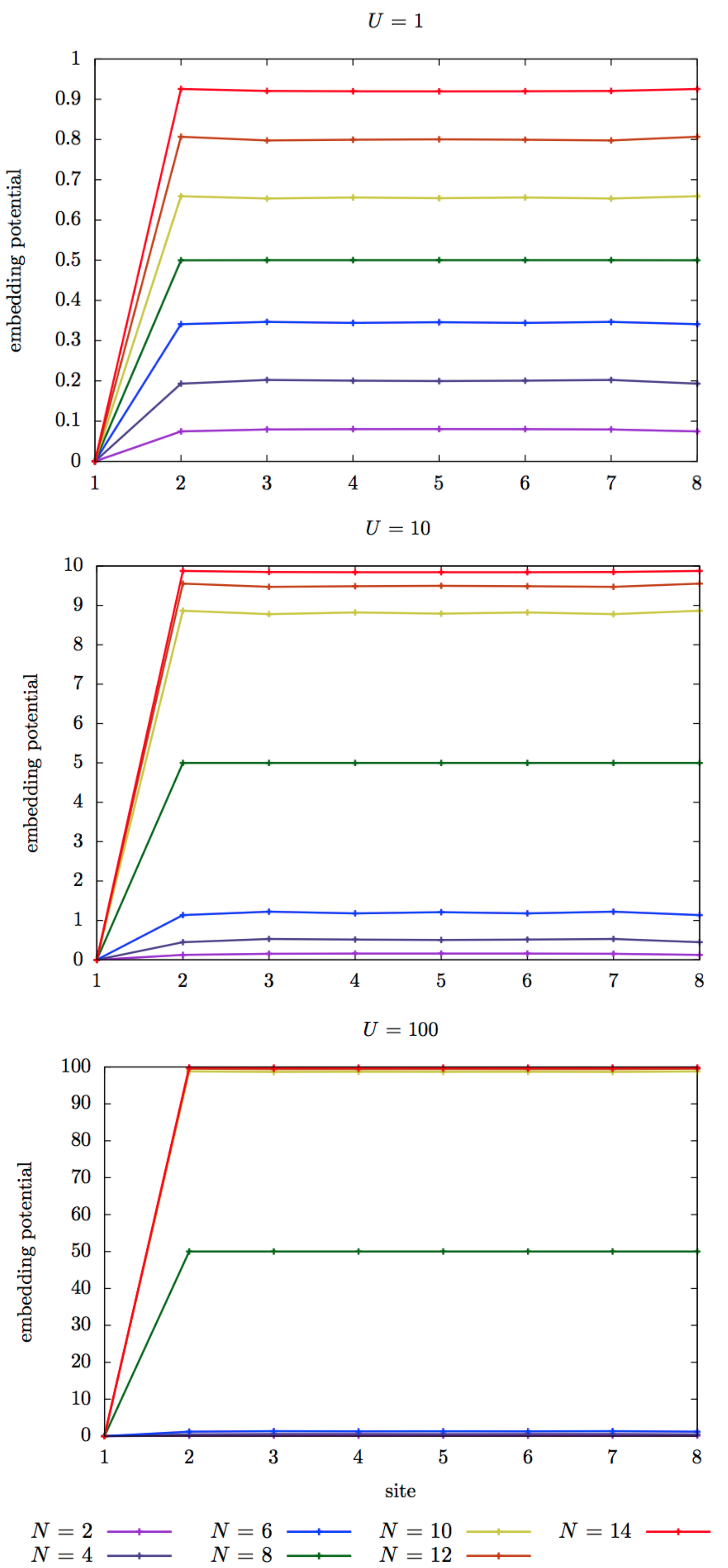}
}
\caption{Exact embedding potentials obtained for uniform 8-site
Hubbard rings with $U = 1$ (top), $U = 10$ (middle) and $U = 100$
(bottom). The site occupations are $N/8$ where $N$ is the number of
electrons. The impurity site is labelled as 1. Potentials are set to zero on the impurity site.}
\label{fig:embedding_pot_LF_transform}
\end{figure}

\begin{figure}
\centering{
\includegraphics[scale=0.4]{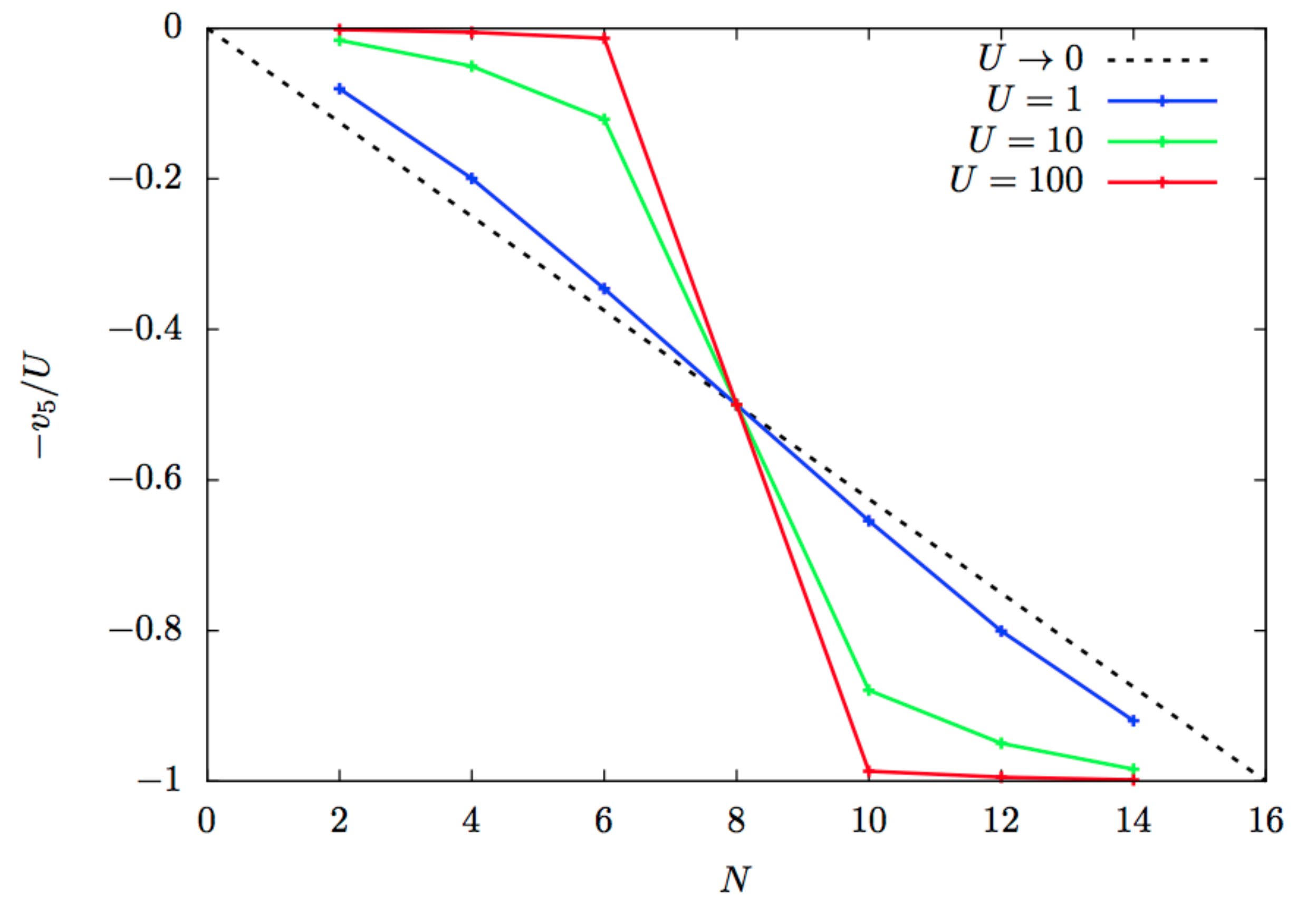}
}
\caption{Exact embedding potentials in the bath (site 5) obtained for uniform 8-site
Hubbard rings with $U = 1$, $U = 10$ and $U = 100$. The site occupations are $N/8$ where $N$ is the number of
electrons. Potentials are set to zero on the impurity site. Comparison
is made with the mean-field approximation ($U\rightarrow0$).}
\label{fig:vLFexact}
\end{figure}

\begin{figure}
\centering{
\includegraphics[scale=0.4]{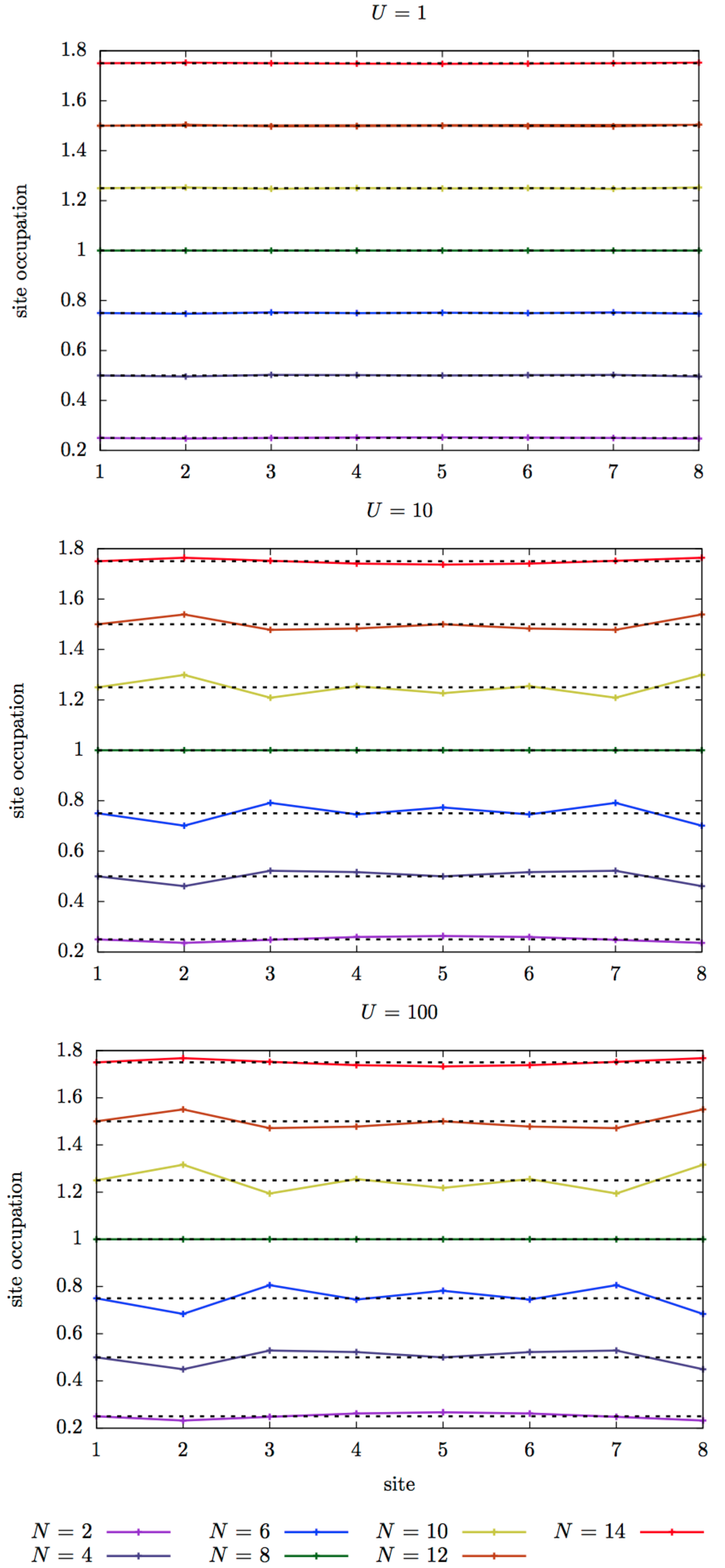}
}
\caption{
site occupations obtained with the impurity-optimised potential for uniform 8-site
Hubbard rings with $U = 1$ (top), $U = 10$ (middle) and $U = 100$
(bottom). $N$ is the number of
electrons. The impurity site is labelled as 1. Dashed lines show uniform
site occupations. See text for further details.
}
\label{fig:LF_transform_imp}
\end{figure}

The so-called $v$-representability of given site occupations~\cite{schindlmayr1995density, DFT_lattice} is usually
considered
for fully-interacting or non-interacting systems. Since SOET relies on
a fictitious embedded impurity system, the problem of what could be
referred to as impurity-interacting $v$-representability arises. In other
words, is it always possible to find a potential such that the embedded
impurity system has the desired site occupations ? This question is
relevant even for uniform site occupations. In the latter case, the
answer is trivial for both non-interacting and fully-interacting systems:
the potential is a constant. This cannot be true anymore if an impurity
is introduced into the system. As shown in
Figure~\ref{fig:embedding_pot_LF_transform}, embedding potentials can be
computed very accurately by means of Legendre--Fenchel transforms for various uniform site occupations in the 8-site
Hubbard ring. These potentials will be referred to as exact embedding potentials in the
following. 
As expected, the difference in potential between the impurity site and
the bath is substantial. Moreover, the potential is not strictly uniform
in the bath, even though the difference in potential between two bath
sites seems to be relatively small. From this observation we conclude
that ILDA is a sound approximation.    
Interestingly, at half-filling ($N = 8$), ILDA becomes exact and the
potential in the bath reduces to its mean-field approximation. Note that
this
feature has already
been observed in the Hubbard dimer (see Figure~\ref{fig:v_vimp}).\\ 
Another important feature of the exact embedding potential is the
discontinuity at half-filling in the strongly correlated limit
($U\rightarrow+\infty$), as highlighted in Figure~\ref{fig:vLFexact}. 
Let us stress that a similar pattern is obtained in the Hubbard dimer
(see Figure~\ref{fig:v_vimp}). In this respect, 2L-ILDA, which relies on
ILDA and the two-level Hubbard model, is also a sound approximation.\\
For analysis purposes, approximate Legendre--Fenchel transforms have been computed
with the potential set to zero in the bath (see Equation~(\ref{LF2})), thus leading to
impurity-optimised embedding potentials. The corresponding site
occupations are shown in Figure~\ref{fig:LF_transform_imp}.
In the weakly correlated regime ($U = 1$), they are almost uniform.
Fluctuations around the exact uniform occupations are shown in Figure~\ref{fig:site-occ-comparison}. 
When $U$ increases, the deviation from uniformity is more visible,
simply because the bath does not contribute to the impurity-optimised
potential. Nevertheless, even for the large $U = 100$ value,
such a potential provides a good starting point for better approximations.
In this respect, ILDA is also relevant in the strongly correlated
regime. Note that the impurity-optimised potential is the best potential
one can get within ILDA since it reproduces the impurity occupation
exactly.

\subsection{Two-level ILDA results}\label{subsec:2L-ILDA_results}

\begin{figure}
\centering{
\includegraphics[scale=0.4]{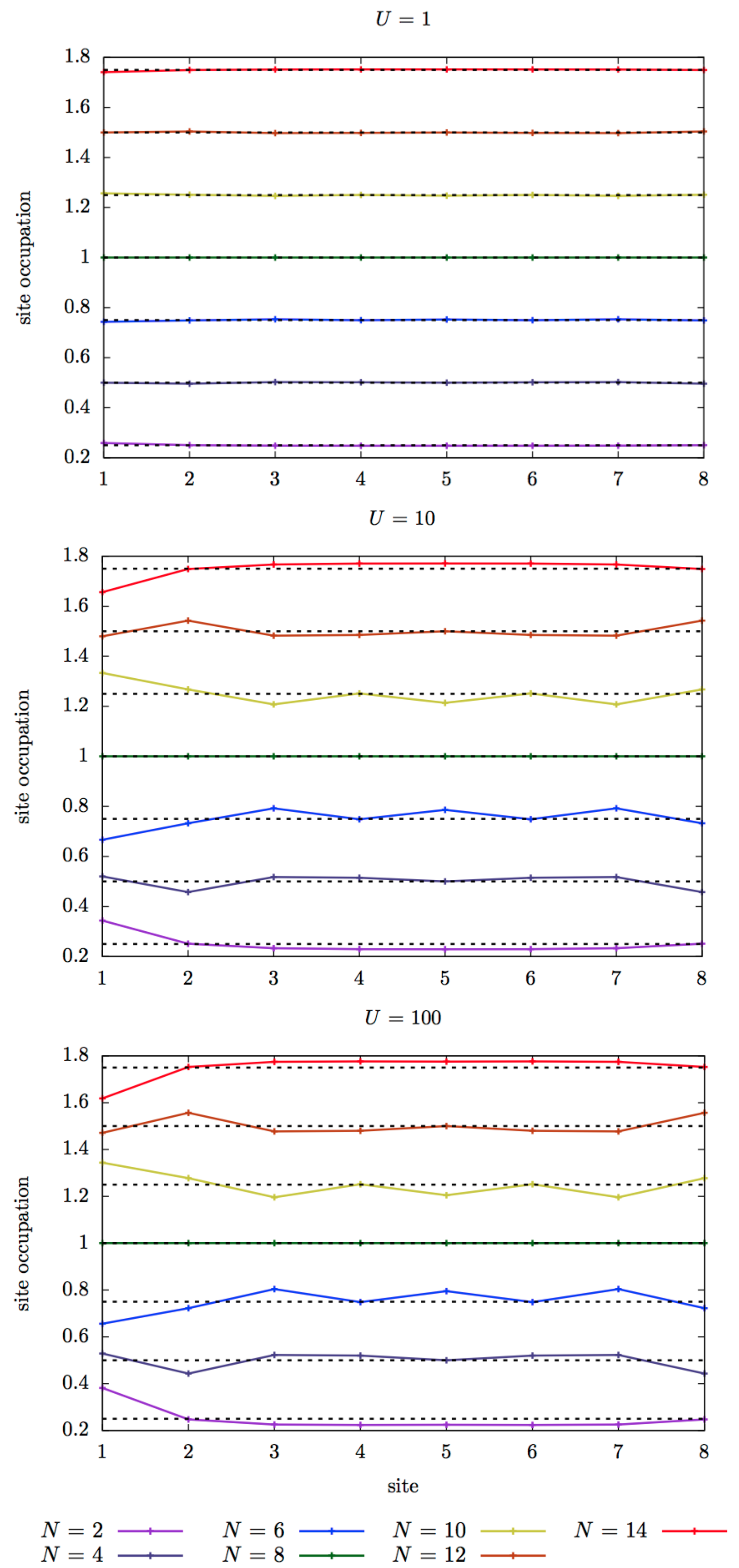}
}
\caption{
site occupations obtained self-consistently with the 2L-ILDA potential for uniform 8-site
Hubbard rings with $U = 1$ (top), $U = 10$ (middle) and $U = 100$
(bottom). $N$ is the number of
electrons. The impurity site is labelled as 1. Dashed lines show uniform
site occupations. See text for further details.
}
\label{fig:ILDA_site_occ}
\end{figure}
\begin{figure}
\centering{
\includegraphics[scale=0.4]{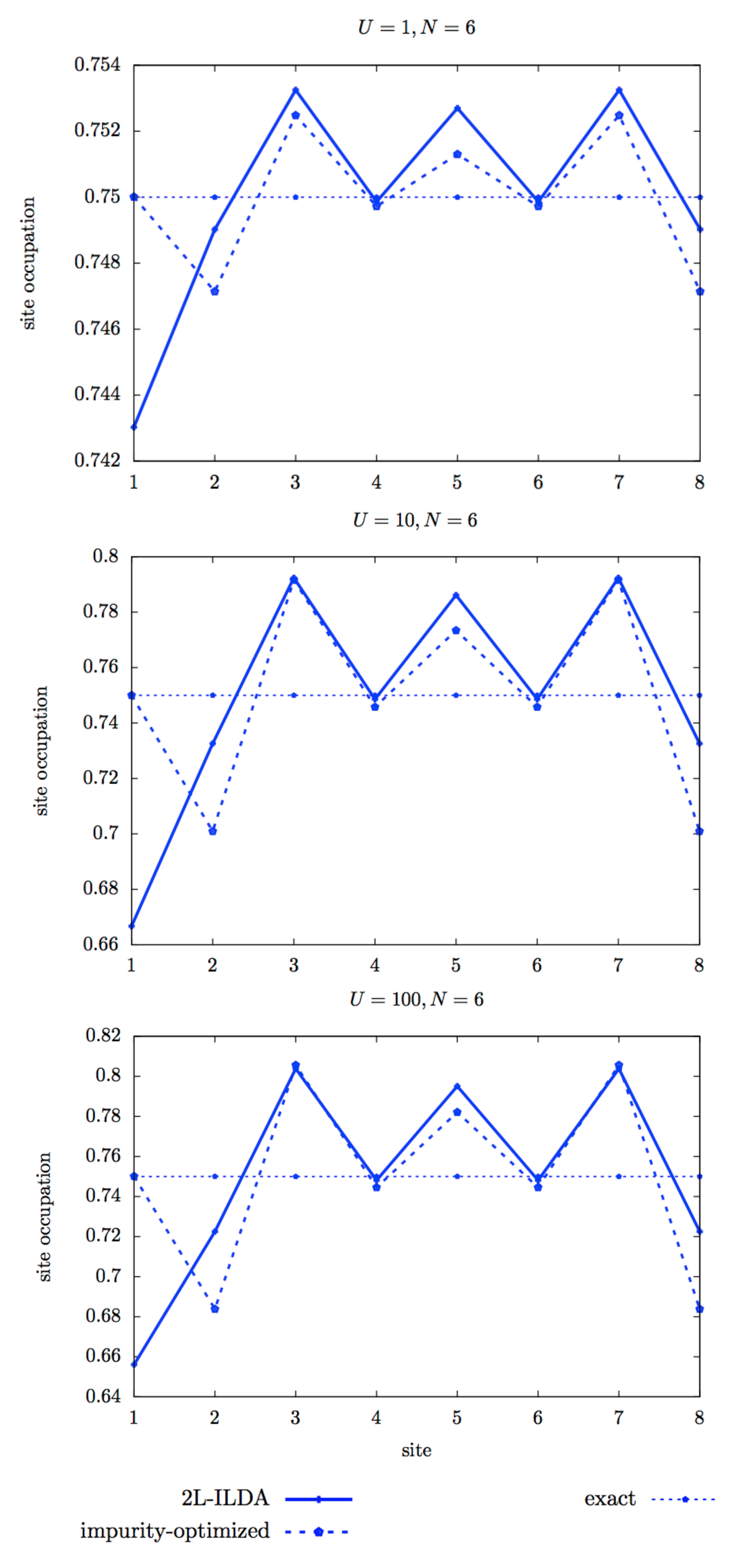}
}
\caption{
site occupations obtained with the exact, impurity-optimised and
self-consistent 2L-ILDA embedding potentials for the uniform 8-site
Hubbard ring with 6 electrons and $U = 1$ (top), $U = 10$ (middle) and $U = 100$
(bottom). The impurity site is labelled as 1.}
\label{fig:site-occ-comparison}
\end{figure}
\begin{figure}
\centering{
\includegraphics[scale=0.4]{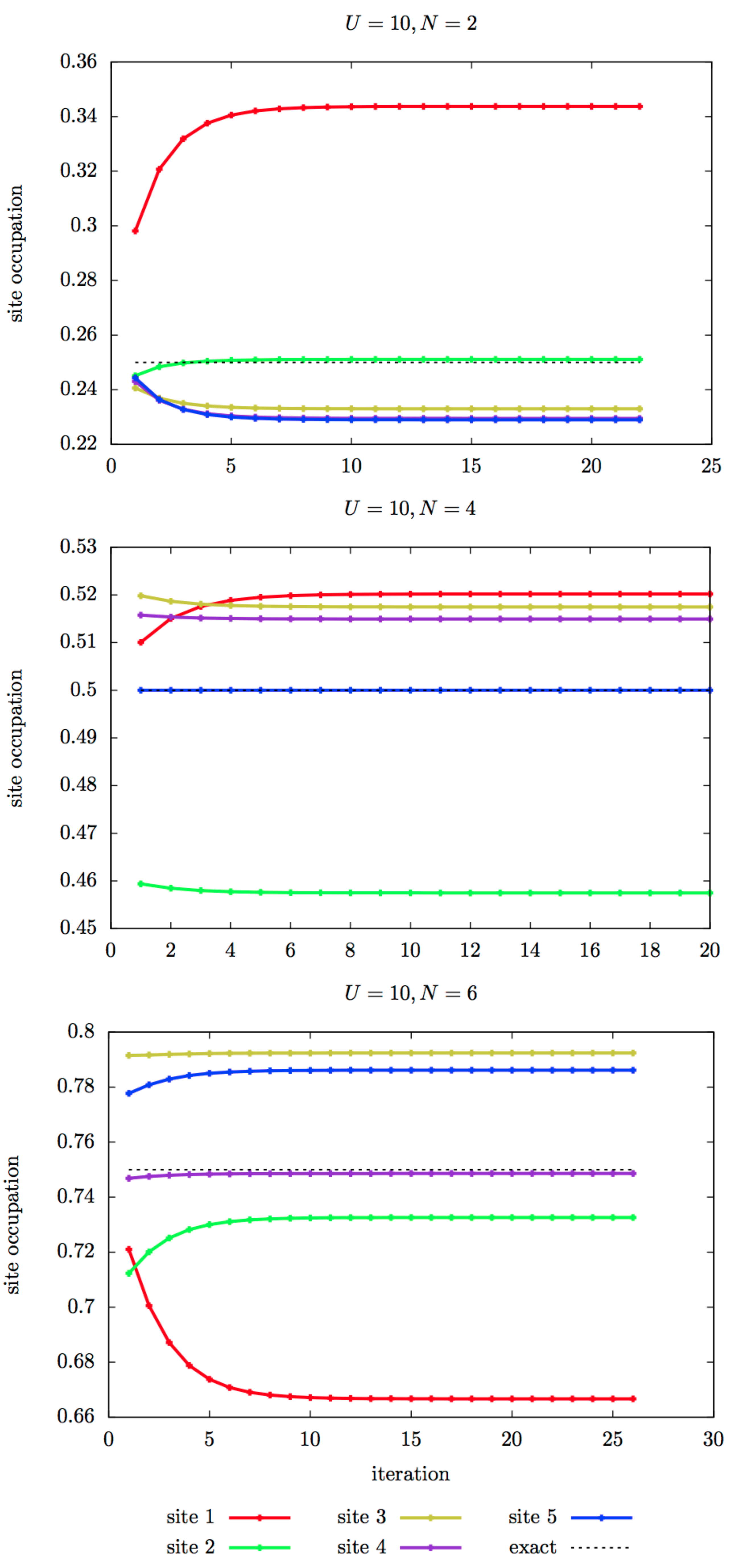}
}
\caption{Convergence of the site occupations in self-consistent 2L-ILDA
calculations for uniform 8-site
Hubbard rings with $N = 2$ (top), $N = 4$ (middle) and $N = 6$
(bottom) electrons. The $U$ parameter is set to 10. The impurity site is
labelled as 1. {Note that, for $N = 4$, the occupation of site 5 is on top of the exact uniform one.}}
\label{fig:occ_it}
\end{figure}
\begin{figure}
\centering{
\includegraphics[width=\textwidth]{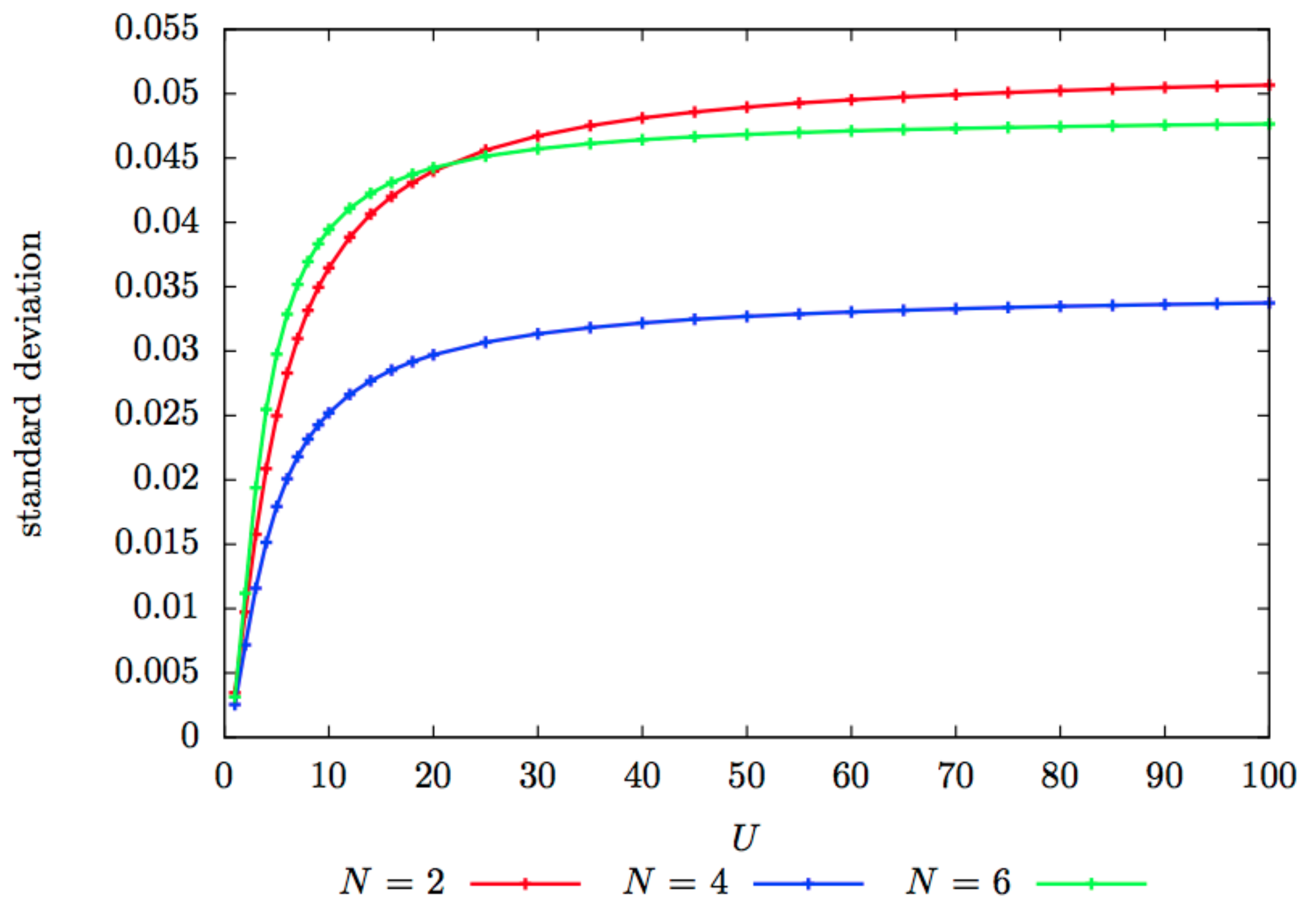}
}
\caption{Standard deviation of converged 2L-ILDA site occupations
obtained when varying $U$ in uniform 8-site
Hubbard rings with $N = 2$, $N = 4$ and $N = 6$ electrons. See text for
further details.}
\label{fig:deviation}
\end{figure}

This section deals with the performance of the 2L-ILDA. Converged site
occupations are shown in Figure~\ref{fig:ILDA_site_occ}. 2L-ILDA is
exact at half-filling ($N=8$), as expected. Indeed, in this case, the potential
is uniform in the bath and equal to its mean-field approximation. Away from half-filling, site occupations deviate from
uniformity. Comparison is made with the 
impurity-optimised potential results in Figure~\ref{fig:site-occ-comparison} for $N = 6$ electrons.
2L-ILDA does not reproduce the correct impurity occupation whereas, by
construction, the impurity-optimised potential scheme does. This was
somehow expected since the site-occupation-functional embedding potential used in
the 2L-ILDA is parametrised for the Hubbard dimer. In the latter case, if the impurity
occupation is $n$, the potential will impose the occupation $2-n$ on the
other site. In other words, we cannot expect a self-consistent 2L-ILDA
calculation to provide strict
uniformity. For analysis purposes, the convergence of the site
occupations is shown in Figure~\ref{fig:occ_it}. Note that it is smooth.
At iteration 0, the impurity occupation is set to the expected uniform
value $N/8$. The site occupations become non-uniform already at the
first iteration in the  
self-consistent calculation of the 2L-ILDA potential.  
The standard deviation from uniformity,
\begin{eqnarray}
\sigma = \sqrt{\sum_{i=1}^L \dfrac{(n_i - N/L)^2}{L}},
\end{eqnarray}
is plotted in Figure~\ref{fig:deviation} with respect to $U$.
It increases monotonically with $U$. The largest slope is obtained in
the range $0 < U < 10$. Then the slope decreases significantly for $10 < U <
30$. In the strongly correlated regime ($U > 30$), the standard
deviation becomes less sensitive to the repulsion strength.\\ 

\begin{figure}
\centering{
\includegraphics[width=\textwidth]{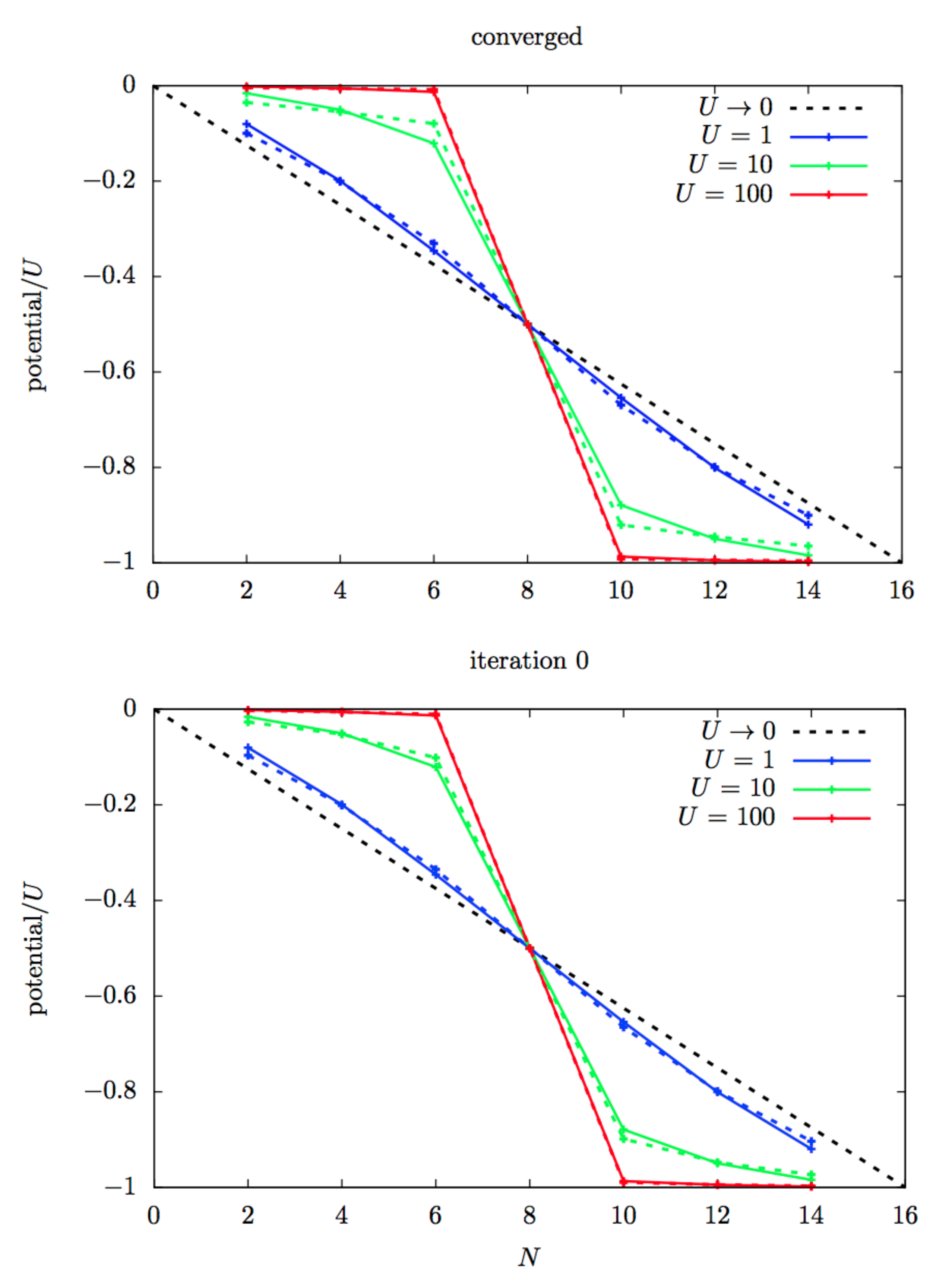}
}
\caption{Impurity-optimised (solid lines) and 2L-ILDA (dashed lines) embedding potentials obtained
for $U=1$, $U=10$ and $U=100$ 
when varying the number $N$ of electrons in a uniform 8-site
Hubbard ring. 
Converged (top) and non-converged (bottom) 2L-ILDA
potentials are shown. The exact (see
Figure~\ref{fig:vLFexact}) and impurity-optimised potentials were found
to be on top of each other. The mean-field approximation ($U\rightarrow 0$) is
also shown for comparison.
}
\label{fig:discontinuity}
\end{figure}

Let us now focus on the 2L-ILDA embedding potential. Comparison is made
with the exact and impurity-optimised potentials in Figure~\ref{fig:discontinuity}.
A nice feature of 2L-ILDA is that the step that leads to a
discontinuity at half-filling in the $U/t\rightarrow+\infty$ limit is relatively well
reproduced. In the weakly correlated regime ($U=1$), self-consistency
does not affect the potential significantly which is relatively close to
the exact one. This statement does not hold for $U = 10$. This can be
related to the deviation from uniformity discussed previously.\\   
\begin{figure}
\centering{
\includegraphics[scale=0.4]{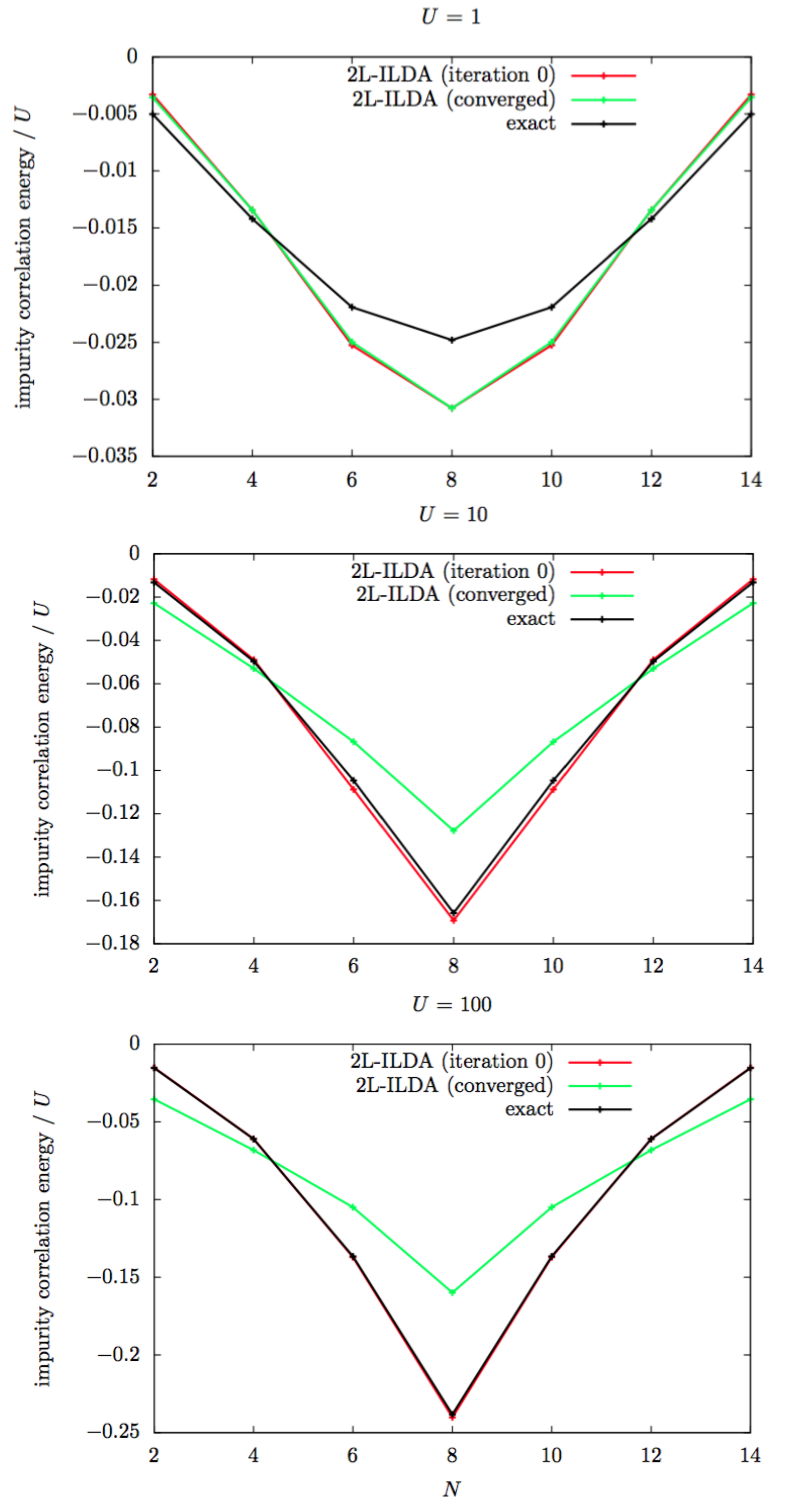}
}
\caption{Exact and 2L-ILDA impurity correlation energies obtained 
for $U=1$ (top), $U=10$ (middle) and $U=100$ (bottom) 
when varying the number $N$ of electrons in a uniform 8-site
Hubbard ring. Converged and non-converged 2L-ILDA results are shown. In
the latter case, the impurity site occupation is set to the expected
exact value $N/8$.}
\label{fig:Ecimp_it0_converged}
\end{figure}

Turning to the impurity correlation energy (see
Figure~\ref{fig:Ecimp_it0_converged}), comparison with the exact and
2L-ILDA (iteration 0) results shows that the error is, like in many
conventional DFT calculations~\cite{kim2013understanding}, essentially
driven by the approximate functional when $U$ is small. This was expected since the impurity correlation
functional used in 2L-ILDA is parametrised for the Hubbard dimer where
the bath reduces to a single site with occupation $2-n$ rather than a
collection of sites which all have the same occupation $n$. On the other hand, 
when the correlation is strong ($U=10$ and $U=100$), the error is
site-occupation-driven. Indeed, at
iteration 0, the impurity occupation is exact and the 2L-ILDA impurity
correlation energy is close to
the exact one. This is probably due to the fact that, in
the strongly correlated limit, interactions become essentially local. As mentioned previously, self-consistency
deteriorates the impurity occupation, thus leading to a substantial error in the
correlation energy. 

\section{Conclusion}\label{sec:conclu}

The extension of DFT to model Hamiltonians such as the Hubbard
Hamiltonian, as well as the formulation of
an exact site-occupation embedding theory (SOET), which was recently proposed by
one of the authors~\cite{fromager2015exact}, have been briefly reviewed. A simple local site-occupation approximation, referred to as 2L-ILDA,
has then been proposed in order to perform practical SOET calculations. It
relies on two approximations: (i) the correlation energy of the
embedded impurity system is assumed to depend on the occupation of the
impurity site 
only and, within such an approximation, (ii) the
embedded impurity correlation functional of the (two-level)
Hubbard dimer is employed. As a proof of concept, the theory has been
tested on a uniform 8-site Hubbard ring. 
Exact and 2L-ILDA calculations have been
performed. In the latter case, the parametrisation of Carrascal \textit{et
al.}~\cite{carrascal2015hubbard} has been used. Promising results were obtained at the 2L-ILDA
level, even though the approximation is quite simple. In the weakly
correlated regime, the error on the impurity correlation energy is
functional-driven while, in the strongly correlated regime, it is
site-occupation-driven. Results might be improved by developing a
functional that depends
explicitly on the impurity nearest neighbours occupations, in the spirit
of the generalised gradient approximation in DFT. The Anderson
model~\cite{Anderson} might be used for developing more accurate ILDA schemes. 
In order to turn SOET into a useful computational method, one should
obviously compute double
occupancies, through the evaluation of energy derivatives with respect
to the on-site repulsion parameter $U$, and perform calculations on
larger rings, for example, 
by combining DMRG~\cite{white1992density} with SOET. 
Note finally that SOET can in principle be applied to quantum chemical
Hamiltonians. As a first step, it might be interesting to 
consider the Richardson Hamiltonian~\cite{richardson1963restricted, richardson1964exact, johnson2013size}.
Work is currently in progress in all these directions.

\section*{Acknowledgements}

The authors are pleased to dedicate this work to Hans J{\o}rgen Aagaard
Jensen on the
occasion of his 60th birthday (congratulations Hans J{\o}rgen and thanks for
everything !). They would like to thank Andreas Savin, Antonin Hommes,
Killian Deur and Laurent Mazouin for
fruitful discussions. E.~Fromager acknowledges financial support from the LABEX
'Chemistry of complex systems' [grant number CSC-VRO-13]
and the ANR (MCFUNEX project [grant number ANR-14-CE06-0014-01]). 
B.~Senjean thanks the Ecole Doctorale des Sciences Chimiques in
Strasbourg for his PhD grant.

\label{lastpage}

\bibliographystyle{tMPH}


\begin{thebibliography}{64}
\providecommand{\url}[1]{\texttt{#1}}
\providecommand{\urlprefix}{URL }
\markboth{E. Fromager}{Molecular Physics}

\bibitem{andersson1992second}
K. Andersson, P.{\AA}. Malmqvist and B.O. Roos,  J. Chem. Phys.  \textbf{96}
  (2), 1218 (1992).

\bibitem{roca2012multiconfiguration}
D. Roca-Sanju{\'a}n, F. Aquilante and R. Lindh,  WIREs Comput Mol Sci
  \textbf{2} (4), 585 (2012).

\bibitem{angeli2001introduction}
C. Angeli, R. Cimiraglia, S. Evangelisti, T. Leininger and J.P. Malrieu,  J.
  Chem. Phys.  \textbf{114} (23), 10252 (2001).

\bibitem{angeli2002n}
C. Angeli, R. Cimiraglia and J.P. Malrieu,  J. Chem. Phys.  \textbf{117} (20),
  9138 (2002).

\bibitem{lyakh2011multireference}
D.I. Lyakh, M. Musia{\l}, V.F. Lotrich and R.J. Bartlett,  Chem. Rev.
  \textbf{112} (1), 182 (2011).

\bibitem{gros1987antiferromagnetic}
C. Gros, R. Joynt and T. Rice,  Phys. Rev. B  \textbf{36} (1), 381 (1987).

\bibitem{sorella2005wave}
S. Sorella,  Phys. Rev. B  \textbf{71} (24), 241103 (2005).

\bibitem{neuscamman2012optimizing}
E. Neuscamman, C. Umrigar and G.K.L. Chan,  Phys. Rev. B  \textbf{85} (4),
  045103 (2012).

\bibitem{zhang1997pairing}
S. Zhang, J. Carlson and J. Gubernatis,  Phys. Rev. Lett.  \textbf{78} (23),
  4486 (1997).

\bibitem{chang2010spin}
C.C. Chang and S. Zhang,  Phys. Rev. Lett.  \textbf{104} (11), 116402 (2010).

\bibitem{yanagisawa2001ground}
T. Yanagisawa, S. Koike and K. Yamaji,  Phys. Rev. B  \textbf{64} (18), 184509
  (2001).

\bibitem{white1992density}
S.R. White,  Phys. Rev. Lett.  \textbf{69}, 2863 (1992).

\bibitem{white1993density}
S.R. White,  Phys. Rev. B  \textbf{48} (14), 10345 (1993).

\bibitem{rodriguez2013multireference}
R. Rodr{\'\i}guez-Guzm{\'a}n, C.A. Jim{\'e}nez-Hoyos, R. Schutski and G.E.
  Scuseria,  Phys. Rev. B  \textbf{87} (23), 235129 (2013).

\bibitem{scuseria2011projected}
G.E. Scuseria, C.A. Jim{\'e}nez-Hoyos, T.M. Henderson, K. Samanta and J.K.
  Ellis,  J. Chem. Phys.  \textbf{135} (12), 124108 (2011).

\bibitem{jimenez2012projected}
C.A. Jim{\'e}nez-Hoyos, T.M. Henderson, T. Tsuchimochi and G.E. Scuseria,  J.
  Chem. Phys.  \textbf{136} (16), 164109 (2012).

\bibitem{rodriguez2012symmetry}
R. Rodr{\'\i}guez-Guzm{\'a}n, K. Schmid, C.A. Jim{\'e}nez-Hoyos and G.E.
  Scuseria,  Phys. Rev. B  \textbf{85} (24), 245130 (2012).

\bibitem{juillet2013exotic}
O. Juillet and R. Fr{\'e}sard,  Phys. Rev. B  \textbf{87} (11), 115136 (2013).

\bibitem{tomita2004many}
N. Tomita,  Phys. Rev. B  \textbf{69} (4), 045110 (2004).

\bibitem{baeriswyl2009variational}
D. Baeriswyl, D. Eichenberger and M. Menteshashvili,  New. J. Phys.
  \textbf{11} (7), 075010 (2009).

\bibitem{lanata2012efficient}
N. Lanata, H.U. Strand, X. Dai and B. Hellsing,  Phys. Rev. B  \textbf{85} (3),
  035133 (2012).

\bibitem{pulay1983localizability}
P. Pulay,  Chem. Phys. Lett.  \textbf{100} (2), 151 (1983).

\bibitem{saebo1993local}
S. Saebo and P. Pulay,  Annu. Rev. Phys. Chem.  \textbf{44} (1), 213 (1993).

\bibitem{hampel1996local}
C. Hampel and H.J. Werner,  J. Chem. Phys.  \textbf{104} (16), 6286 (1996).

\bibitem{tsuchimochi2015density}
T. Tsuchimochi, M. Welborn and T. Van~Voorhis,  J. Chem. Phys.  \textbf{143}
  (2), 024107 (2015).

\bibitem{DMFT_correlation_limit}
A. Georges, G. Kotliar, W. Krauth and M.J. Rozenberg,  Rev. Mod. Phys.
  \textbf{68} (1996).

\bibitem{kotliar2004strongly}
G. Kotliar and D. Vollhardt,  Phys. Today  \textbf{57} (3), 53 (2004).

\bibitem{DMFT_calculations}
G. Kotliar, S.Y. Savrasov, K. Haule, V.S. Oudovenko, O. Parcollet and C.A.
  Marianetti,  Rev. Mod. Phys.  \textbf{78}, 865 (2006).

\bibitem{held2007electronic}
K. Held,  Adv. Phys.  \textbf{56} (6), 829 (2007).

\bibitem{DMFT_quantum}
D. Zgid and G.K.L. Chan,  J. Chem. Phys.  \textbf{134} (2011).

\bibitem{kananenka2015systematically}
A.A. Kananenka, E. Gull and D. Zgid,  Phys. Rev. B  \textbf{91} (12), 121111
  (2015).

\bibitem{lan2015communication}
T.N. Lan, A.A. Kananenka and D. Zgid,  J. Chem. Phys.  \textbf{143} (24),
  241102 (2015).

\bibitem{dahlen2005self}
N.E. Dahlen and R. van Leeuwen,  J. Chem. Phys.  \textbf{122} (16), 164102
  (2005).

\bibitem{phillips2014communication}
J.J. Phillips and D. Zgid,  J. Chem. Phys.  \textbf{140} (24), 241101 (2014).

\bibitem{DMET_vs_DMFT}
G. Knizia and G.K.L. Chan,  Phys. Rev. Lett.  \textbf{109}, 186404 (2012).

\bibitem{zheng2016ground}
B.X. Zheng and G.K.L. Chan,  Phys. Rev. B  \textbf{93} (3), 035126 (2016).

\bibitem{bulik2014density}
I.W. Bulik, G.E. Scuseria and J. Dukelsky,  Phys. Rev. B  \textbf{89}, 035140
  (2014).

\bibitem{nakamura1959two}
K.i. Nakamura,  Prog. Theor. Phys.  \textbf{21} (5), 713 (1959).

\bibitem{kutzelnigg1964direct}
W. Kutzelnigg,  J. Chem. Phys.  \textbf{40} (12), 3640 (1964).

\bibitem{coleman1965structure}
A. Coleman,  J. Math. Phys.  \textbf{6} (9), 1425 (1965).

\bibitem{fromager2015exact}
E. Fromager,  Mol. Phys.  \textbf{113} (5), 419 (2015).

\bibitem{gunnarsson1986density}
O. Gunnarsson and K. Sch{\"o}nhammer,  Phys. Rev. Lett.  \textbf{56} (18), 1968
  (1986).

\bibitem{DFT_ModelHamiltonians}
K. Capelle and V.L. {Campo Jr.},  Phys. Rep.  \textbf{528}, 91 (2013).

\bibitem{DFT_lattice}
K. Schonhammer, O. Gunnarsson and R. Noack,  Phys. Rev. B  \textbf{52}, 2054
  (1995).

\bibitem{LDA_Luttinger}
K. Capelle, N.A. Lima, M.F. Silva and L.N. Oliveira,  Phys. Rev. Lett.
  \textbf{90}, 146402 (2003).

\bibitem{carrascal2015hubbard}
D.J. Carrascal, J. Ferrer, J.C. Smith and K. Burke,  J. Phys. Condens. Matter
  \textbf{27} (39), 393001 (2015).

\bibitem{anisimov1991band}
V.I. Anisimov, J. Zaanen and O.K. Andersen,  Phys. Rev. B  \textbf{44}, 943
  (1991).

\bibitem{liechtenstein1995density}
A. Liechtenstein, V. Anisimov and J. Zaanen,  Physical Review B  \textbf{52}
  (8), R5467 (1995).

\bibitem{savinbook}
A. Savin, \emph{Recent Developments and Applications of Modern Density
  Functional Theory}   (Elsevier, Amsterdam, 1996), p. 327.

\bibitem{leininger1997combining}
T. Leininger, H. Stoll, H.J. Werner and A. Savin,  Chem. Phys. Lett.
  \textbf{275} (3), 151 (1997).

\bibitem{pollet2002combining}
R. Pollet, A. Savin, T. Leininger and H. Stoll,  J. Chem. Phys.  \textbf{116}
  (4), 1250 (2002).

\bibitem{srDFT}
J. Toulouse, F. Colonna and A. Savin,  Phys. Rev. A  \textbf{70}, 062505
  (2004).

\bibitem{chayes1985density}
J. Chayes, L. Chayes and M.B. Ruskai,  J. Stat. Phys.  \textbf{38} (3-4), 497
  (1985).

\bibitem{LFTransform-Lieb}
E.H. Lieb,  Int. J. Quantum Chem.  \textbf{24} (3), 243 (1983).

\bibitem{bergfield2012bethe}
J.P. Bergfield, Z.F. Liu, K. Burke and C.A. Stafford,  Phys. Rev. Lett.
  \textbf{108} (6), 066801 (2012).

\bibitem{liu2012accuracy}
Z.F. Liu, J.P. Bergfield, K. Burke and C.A. Stafford,  Phys. Rev. B
  \textbf{85} (15), 155117 (2012).

\bibitem{liu2015coulomb}
Z.F. Liu and K. Burke,  Phys. Rev. B  \textbf{91} (24), 245158 (2015).

\bibitem{lanczos}
E. Dagotto,  Rev. Mod. Phys.  \textbf{66}, 763 (1994).

\bibitem{schindlmayr1995density}
A. Schindlmayr and R. Godby,  Phys. Rev. B  \textbf{51} (16), 10427 (1995).

\bibitem{kim2013understanding}
M.C. Kim, E. Sim and K. Burke,  Phys. Rev. Lett.  \textbf{111} (7), 073003
  (2013).

\bibitem{Anderson}
P.W. Anderson,  Phys. Rev.  \textbf{124}, 1 (1961).

\bibitem{richardson1963restricted}
R. Richardson,  Phys. Lett.  \textbf{3} (6), 277 (1963).

\bibitem{richardson1964exact}
R. Richardson and N. Sherman,  Nucl. Phys.  \textbf{52}, 221 (1964).

\bibitem{johnson2013size}
P.A. Johnson, P.W. Ayers, P.A. Limacher, S. De~Baerdemacker, D. Van~Neck and P.
  Bultinck,  Comp. Theor. Chem.  \textbf{1003}, 101 (2013).

\end{thebibliography}

\newcommand{\Aa}[0]{Aa}

\end{document}